\documentclass[useAMS,usenatbib]{mn2e}
\usepackage{graphicx}
\graphicspath{{images_de_larticle/}}

\usepackage{amsmath}
\usepackage{placeins}

\usepackage{float}
\usepackage{stfloats}

\usepackage[usenames,dvipsnames]{color}
\usepackage[normalem]{ulem}

\def\lya{Ly$\alpha$~}

\usepackage{lipsum}
\usepackage[bottom,norule,hang]{footmisc}
\title[Calibrating  cosmological radiative transfer  simulations with \lya forest data]{Calibrating
  cosmological radiative transfer  simulations with \lya forest data: 
   Evidence for large spatial  UV background fluctuations at $z\sim 5.6-5.8$ due to rare bright sources}
\author[J. Chardin et al.]{Jonathan
  Chardin$^{1}$\thanks{E-mail:jc@ast.cam.ac.uk}, 
Martin G. Haehnelt$^{1}$, Dominique Aubert$^{2}$ and Ewald Puchwein$^{1}$\\
$^{1}$Kavli Institute for Cosmology and Institute of Astronomy, Madingley Road, Cambridge CB3 0HA\\
$^{2}$Observatoire Astronomique de Strasbourg, Universit\'e de Strasbourg, CNRS UMR 7550, 11 rue de l'Universit\'e, F-67000 Strasbourg, France\\}

\begin{document}


\date{Accepted / Received }

\pagerange{\pageref{firstpage}--\pageref{lastpage}} \pubyear{2014}

\maketitle

\begin{abstract}

We calibrate  here cosmological radiative transfer simulations with ATON/RAMSES with a range of  measurements  of the \lya opacity from QSO absorption spectra.  
We find the  \lya opacity to be  very sensitive  to the exact timing of hydrogen reionisation.  
Models reproducing the measured evolution of the mean photoionisation rate and average mean free path 
reach overlap at $z\sim 7$ and predict an accelerated evolution of the \lya
opacity at $z>6$ consistent  with the rapidly evolving luminosity function  of \lya emitters in this redshift range.  
Similar to "optically thin" simulations our full radiative transfer simulations fail, however, 
to  reproduce the  high-opacity tail of the \lya opacity PDF at $z>5$. We argue that this is due 
to spatial UV fluctuations in the post-overlap phase of reionisation on substantially larger scales than predicted by our 
source model, where the ionising emissivity is dominated by large numbers of sub-$\mathrm{L_*}$ galaxies.
We further argue that this suggests a significant contribution to the ionising UV background by much rarer bright sources
at high redshift.

\end{abstract}



\section{Introduction}
\label{intro}

The $\mathrm{Ly}\alpha$ forest is a valuable probe of the underlying matter density field, as well as of the 
temperature and ionisation state of the Inter-Galactic-Medium (IGM) at high redshift 
(see \citealt{1998ARA&A..36..267R}, \citealt{2009RvMP...81.1405M} for reviews). 
There is strong evidence that, despite a significantly increased effective 
\lya optical depth, even the   highest redshift QSOs at $z\sim 6$ still probe  the 
post-reionisation highly-ionised IGM (\citealt{2007MNRAS.382..325B},    
{\it cf.} \citealt{2006ApJ...639L..47L} and \citealt{2010MNRAS.407.1328M}),  
apart from perhaps the notable  $z=7.085$ QSO ULAS J1120+0641 which appears to 
show signs of a red damping wing suggesting a volume fraction of neutral hydrogen of 10\% or more 
(\citealt{2006AJ....132..117F}, \citealt{2011Natur.474..616M} and \citealt{2011MNRAS.416L..70B}), 
but see Bosman \& Becker (2015) (in prep) for a recent reassessment of the significance of the weak \lya emission 
of ULAS J1120+0641.

This is in good agreement with the constraints on the reionisation history by the  CMB which 
also suggests that most of hydrogen in the Universe has been reionised   earlier than $z\sim 6$,
even though the most recent Planck (\citealt{2015arXiv150201589P}) measurements
suggest a somewhat later (end of) reionisation than earlier CMB measurements.

The accelerating  evolution of the \lya optical depth at $z>5$ (\citealt{2006AJ....132..117F}) and the apparent 
rapid decrease of the observed \lya emission of high-redshift galaxies (\citealt{2013MNRAS.429.1695B} and \citealt{2014MNRAS.440.3309D}) also  suggest 
that at $z\sim6$ we are witnessing the final stages of the reionisation process 
(\citealt{2010ApJ...723..869O}, \citealt{2011ApJ...734..119K}, \citealt{2011ApJ...743..132P}, 
\citealt{2013ApJ...775L..29T}, \citealt{2014MNRAS.443.2831C}, \citealt{2014ApJ...788...87F}, \citealt{2014ApJ...793..113P}, \citealt{2014arXiv1404.6066K} 
and \citealt{2014arXiv1412.4790C}). 
This has been further corroborated by a recent accurate measurement of \lya optical depth PDF at $4<z<6$ by \citet{2015MNRAS.447.3402B} based on a large sample of 
high-quality high-redshift QSO absorption spectra.  

Simulation of the hydrogen \lya forest at $z<5$ have mostly been performed 
assuming that  hydrogen is already highly ionised, that the Universe can be assumed to be optically thin for hydrogen ionising radiation  and that radiative transfer effects can be neglected. The UV background in these  simulations is assumed to be spatially homogeneous as in widely used  spectral synthesis models of the UV background
(\citealt{1996ApJ...461...20H}, \citealt{2001cghr.confE..64H}, \citealt{2012ApJ...746..125H}).

Such simulations have been used  widely to infer the hydrogen photoionisation rate by rescaling the optical depth in the simulations such that it
matches that observed  (\citealt{1997ApJ...489....7R}, \citealt{1998MNRAS.301..478T}, \citealt{2005MNRAS.357.1178B}, \citealt{2007MNRAS.382..325B}, \citealt{2008ApJ...682L...9F}, 
\citealt{2008ApJ...688...85F}, \citealt{2011MNRAS.412.2543C} and \citealt{2013MNRAS.436.1023B}).
When comparing  such simulations to their measurements of the 
\lya optical depth PDF  \citet{2015MNRAS.447.3402B} reported rapidly increasing  deviations for the high opacity tail of the 
PDF at $z>5$ and interpreted that being due to spatial fluctuations in the 
photoionisation rate in the post-overlap phase   of reionisation as expected in models for inhomogeneous reionisation 
(\citealt{2000ApJ...530....1M}).  

More detailed modelling of this cleary requires full cosmological radiative
transfer simulations. Most cosmological radiative  transfer simulations have concentrated on the 
large scale topology of reionisation (\citealt{2001NewA....6..437G}, \citealt{2002ApJ...572..695R}, \citealt{2006MNRAS.369.1625I}, \citealt{2007ApJ...670...39L},  
\citealt{2007ApJ...671....1T}, \citealt{2008MNRAS.387..295A}, \citealt{2009MNRAS.400.1049F}, \citealt{2009MNRAS.396.1383P}, \citealt{2010ApJ...724..244A},
\citealt{2012A&A...548A...9C}) in rather large simulation boxes. 
Fully understanding the growth of ionised regions  requires, however, not only an accurate modelling of the large scale distribution 
of the sources  of ionising radiation but also of the sinks of ionising radiation as is being increasingly realised 
(\citealt{2000ApJ...530....1M}, \citealt{2005MNRAS.363.1031F}, \citealt{2009MNRAS.394.1667F}, \citealt{2009MNRAS.394..960C}, 
\citealt{2011ApJ...743...82M}, \citealt{2013ApJ...771...35K}, \citealt{2014ApJ...793...29G}, 
\citealt{2014ApJ...793...30G}, \citealt{2014arXiv1412.4790C} and \citealt{2015MNRAS.446..566M}).

This makes accurate cosmological radiative transfer simulations numerically 
extremely demanding in terms of dynamic range and hybrid techniques to treat small and large scale separately  are being developed
(\citealt{2007ApJ...669..663M}, \citealt{2014arXiv1412.4790C} and \citealt{2015MNRAS.446..566M}).

Accurate modelling of the sinks of ionising radiation is particularly important for modelling of \lya forest data  and this 
is the main focus  of the paper here (see also \citealt{2009MNRAS.396.1383P}). 
For this, we use here a suite of rather high-resolution 
hydro-dynamical RAMSES simulations (\citealt{2002A&A...385..337T}) post-processed with the GPU-based radiative transfer code ATON (\citealt{2008MNRAS.387..295A}). 
Note that the box size of our simulation are smaller than those of many other  reionisation simulations in order to properly resolve 
regions optically thick to ionising radiation in the post-overlap phase.

The paper is organized as follows: In section \ref{Simu}, we present the main features of the simulations performed in the context of this work.
In section \ref{results}, we present our results first for the evolution of global quantities in the 
simulation and second for the evolution of Lyman-alpha forest statistics in our different model.
We discuss our results and make predictions for the evolution of the ionising  UV background  in section \ref{IGM_prop} and we 
give our conclusions  and present prospects concerning further work in section \ref{prospects}.
Our simulations  assume the following cosmological parameter (as derived from the 2013 Planck temperature power spectrum data alone, \citealt{2014A&A...571A..16P}) : 
$\Omega_m = 0.3175$, $\Omega_{\Lambda} = 0.6825$,
$\Omega_b = 0.048$, $h = 0.6711$, $\sigma_8 = 0.8344$, and $n_s = 0.9624$.

\section{Numerical simulations}
\label{Simu}

We briefly describe here our  modelling of the spatial distribution of matter  with cosmological 
hydro-simulations with the code RAMSES  and the radiative transfer calculation performed as a post-processing step 
on the hydrodynamic  simulation  outputs with the code ATON.  We will take particular care to model  
the \lya forest in QSO absorption spectra during the post-overlap phase of reionization. We 
benchmark our modelling against the widely used  (homogeneous) UV background model of \citet{2012ApJ...746..125H} (HM2012 hereafter),
which takes into account  a broad set of recent data.

\subsection{Cosmological hydrosimulations with RAMSES}
\label{ramsessimulation}

We have performed cosmological hydro-simulations with the adaptive mesh refinement code RAMSES (\citealt{2002A&A...385..337T}). 
In RAMSES  the  gas dynamics is solved with a 2nd order Godunov scheme  combined with a HLLC Riemann solver. 
The collisionless dynamics of the dark matter (DM) is represented by DM particles and the  
gravitational potential is calculated with  a multi-resolution multi-grid solver. 
Initial conditions were produced with MUSIC (\citealt{2013ascl.soft11011H}).
 
Star formation is included following the prescription of \citet{2006A&A...445....1R}.
The star formation recipe assumes that above  a baryon over-density of 
$\delta \sim 50$   gas transforms into stars of constant mass  with  efficiency $\epsilon = 0.01$ per free fall time.
We do not include stellar or AGN feedback and metal enrichment has also not been included 
(see \citealt{2015arXiv150300734B} for a recent cosmological radiative transfer simulation 
based on the Illustris simulation with AGN and stellar feedback tuned to form a realistic galaxy population).

The usual primordial cooling processes for hydrogen and helium are taken into account: collisional ionisation cooling,
recombination cooling, dielectronic recombination cooling, collisional excitation cooling, 
bremsstrahlung and inverse Compton cooling (see \citealt{1998MNRAS.301..478T}).
Simulation  outputs are generated every 40 Myrs from $z\sim 30$ to $z\sim2$ which results in  80 outputs for 
each of the simulations described further in section \ref{Simulation_set}.

\subsection{Radiative transfer in post-processing with ATON}
\label{RT}

The radiative transfer calculations have been  performed  as a  post-processing step with  the ATON code. 
ATON is described in \citet{ 2008MNRAS.387..295A} and has been mainly used to model hydrogen reionisation  (\citealt{2010ApJ...724..244A}, 
\citealt{2012A&A...548A...9C} and \citealt{2014A&A...568A..52C}) and in particular 
the reionisation of our Local Goup of galaxies (\citealt{2013ApJ...777...51O} and \citealt{2014ApJ...794...20O}).
ATON employs  a moment-based description of the radiative transfer equation, based on the M1 approximation that provides a simple 
and local closure relation between radiative pressure and radiative energy density. The code takes advantage of GPU acceleration to solve the 
equations explicitly in conservative form  while satisfying a very strict Courant condition, i.e. $\Delta t < c \, \Delta x / 3$,
where $\Delta t$ is the time step, $\Delta x$ the cell size and $c$ the speed of light.

Note that the radiative transfer in ATON uses the actual speed of light $c$ and not a reduced speed of light approximation 
(see \citealt{2001NewA....6..437G} and \citealt{2015arXiv150300734B}).
Note further that we have not used here ATON's ability to follow photons with a  range of frequencies.  We have used instead  
the computationally cheaper options of tracing only monochromatic ionising photons  with  energy of 20.27 eV, 
approximately  the mean photon energy of  a 50000 K black-body spectrum   (see also \citealt{2009A&A...495..389B}). 

The radiative  transfer is performed on a  grid with resolution equal to that of the coarse base grid of our RAMSES simulations. 
Note that ATON is not able to follow the AMR structure of RAMSES  and that the  RAMSES simulations used 
do not include any level of mesh refinement.
Note further that recently a version of this radiative transfer scheme has been directly implemented in RAMSES (\citealt{2013MNRAS.436.2188R}) 
that couples self-consistently radiative forces to the  the dynamics of the gas.  This version  is,  however, computationally 
much more expensive and  we do not use it here. 
During  post-processing the density and temperature fields from the RAMSES simulations are updated 
within the radiative transfer calculation every 40 Myrs corresponding to the frequency 
of  hydrodynamic simulation  outputs obtained with RAMSES.

\subsection{Modelling of ionising sources}
\label{photheating}

\subsubsection{Optically thin simulations with the HM2012 UV background model}

Extensive previous modelling of \lya forest data by some of the authors was based on hydrodynamical simulations 
with P-GADGET3 (an improved version of GADGET2 last described in \citealt{2005MNRAS.364.1105S}) neglecting radiative transfer. 
To make contact to this work we have first compared  ``optically thin" RAMSES 
simulations with a homogeneous UV background  with comparable P-GADGET3 simulations to make sure that we obtain consistent results.

For this purpose, we have implemented the latest  version of the UV background of 
HM2012 in the RAMSES code. In the optically thin approximation at any given time 
 the photoionisation  and photoheating rates  are independent of location (and density).
Every part of the simulation sees the same UV background intensity. The time evolution of the space averaged 
UV background intensity  is calculated  by solving a global radiative transfer equation  with an empirical opacity model and a source function 
based on the   observed UV luminosity function of quasar and star forming galaxies.
The HM2012 background model calculated in this way should give a reasonable approximation 
to the spatially averaged photoionisation rates, especially once the Universe is fully ionised 
(see \citealt{2015arXiv1410.1531P} for a recent discussion).
By construction, it will however not be able to describe the growth of individual HII regions 
before overlap and the persistence of spatial amplitude fluctuation in the UV flux and therefore 
photoionisation rate for some time after overlap. 

We will later compare  such optically thin simulations to our full cosmological radiative transfer 
simulation with ATON where the ionising radiation is propagated from discrete sources 
into the surrounding IGM. Note that we do not try to model the ionising radiation from star formation in our simulation 
self-consistently (other than, e.g. \citealt{2015arXiv150300734B}), but that we assume an ionising volume emissivity scaled to that assumed 
in the  HM2012  UV background model and identify the dark matter haloes  in our simulations as the ionising sources.
The dark matter haloes are identified with the HOP halo finder (\citealt{1998ApJ...498..137E}) with a minimal halo mass consisting of 10 particles, 
corresponding to the minimal masses reported in Tables \ref{tab1} and \ref{tab2}. 
We assume  that dark matter haloes above our mass thresholds act as ionising sources with  emissivities
proportional to  halo  mass $M$   similar to the assumptions in  \citet{2006MNRAS.369.1625I},
\begin{equation}
\mathrm{\dot{N}_{\gamma} (z) =\alpha(z) M}.
\label{source_halo}
\end{equation}
\noindent

Note here that simple models where the UV luminosity is proportional to halo mass
appear to fit (UV) luminosity functions of high-redshift galaxies remarkably well 
(see \citealt{2010ApJ...714L.202T}).
We will investigate this further in Appendix \ref{LFevolution}.

\subsubsection{Choosing the ionising emissivity for the ATON simulations}
\label{calibration_emissivity}

\begin{figure*}
   \begin{center}
    \begin{tabular}{cc}
      \includegraphics[width=8cm,height=7cm]{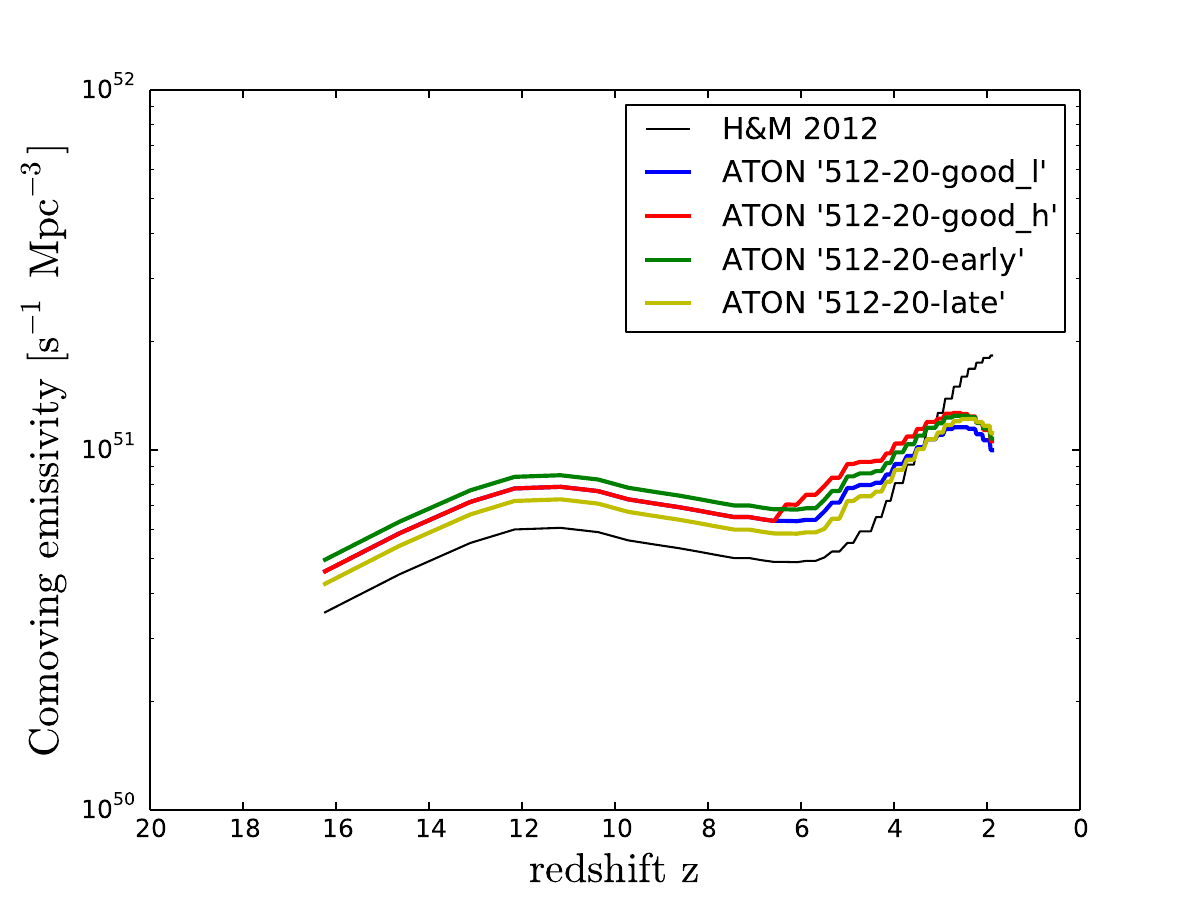} &
      \includegraphics[width=8cm,height=7cm]{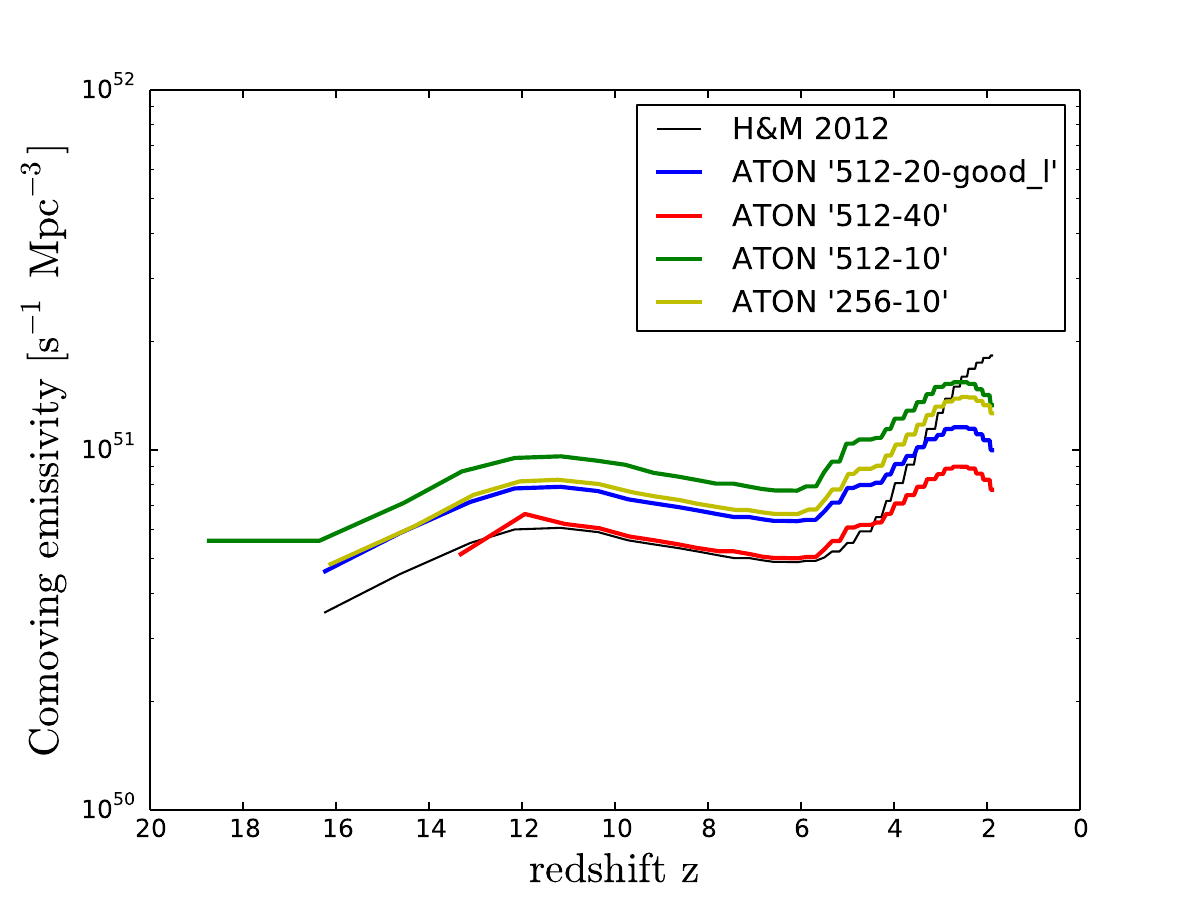} \\
      (a)  & (b)\\
\end{tabular}    
\caption{(a): Evolution of the comoving emissivity of ionising photons (in $\mathrm{s^{-1}Mpc^{-3}}$).
The black solid curve  represents the evolution of the emissivity from HM2012 integrated over all 
frequencies of ionising photons. The other curves  shows the  ionising  emissivities for the different ionisation histories 
studied with our  ATON simulations. The emissivities have  been re-scaled  
relative to the HM2012  model  to  reproduce observed photoionisation rates and to account for the resolution dependent
recombinations in the host haloes of  ionising sources.    (b): same as (a) but for the simulations with 
different box sizes and resolution as described in the text.}
    \label{emissivity_calibration_1}
  \end{center}
 \end{figure*}

In order to obtain a realistic post-overlap  \lya forest and for an easier comparison with our 
optically thin simulations  we have scaled our emissivities in the ATON simulations such that the volume average are close to that of  the HM2012 model,
shown as the black solid curve in Fig.~\ref{emissivity_calibration_1}. The required re-scaling 
will be discussed in detail below.

Note again that we thereby  have not tried  to match the space density of sources in the 
HM2012 model in our ATON simulations, but just the redshift evolution of the integrated emissivity. 
For a given population of DM haloes 
identified as ionising sources this fixes then the evolution of the efficiency parameter $\alpha(z)$ of equation \ref{source_halo}.
The HM2012 model uses a  simple escape fraction to model the fraction of ionising photons escaping 
galaxies,
\begin{equation}
 \left<\mathrm{f_{\rm esc}}\right> = 1.8 \times 10^{-4}(1 + z)^{3.4}.
\end{equation}

As ATON is modelling the escape of photons from the source positions explicitly we have to account for this and boost the escaping ionising emissivity 
compared to ``optically thin" simulations with a spatial homogeneous UV background model like  HM2012  
by a resolution dependent factor that accounts for the recombinations in the host haloes of our ionising sources. 
As we will see later  ``recombination"  boost factors  in the range $\sim 1.05-2$ gave reasonable reionisation histories.  
Note that these boost factors should be equal to the inverse of the escape fraction in the HM2012 background model if our simulations 
were able to correctly model  the escape of ionising photons from real galaxies.  
That these factors are significantly smaller  just means that they are clearly and not surprisingly not  able to do this.   
As we will see later the redshift evolution of the ionising emissivity of  HM2012 does not get  the rather flat evolution of the photoionisation rate at $6>z>2$ as probed by \lya forest data quite right.  We were able to reproduce the \lya forest data 
and the inferred photoionisation rates by scaling the HM2012 ionising emissivities with a redshift dependent scaling factor as follows,

\begin{equation}
b(z) = \left\lbrace
\begin{array}{ccc}
 a ={\rm const} & \mbox{if} & z\ge z_1\\
& & \\
a \left[\frac{z}{z_1}\right]^{\alpha_1} & \mbox{if} & z_2<z\le z_1\\
& & \\
a \left[\frac{z_2}{z_1}\right]^{\alpha_1} \left[\frac{z}{z_2}\right]^{\alpha_2}  & \mbox{if} & z\le z_2\\
\end{array}\right.
\label{equpowrlaw}
\end{equation}
with suitable values of $a$, $z_1$, $z_2$, $\alpha_1$ and $\alpha_2$ as shown in Tables \ref{tab1} and \ref{tab2}.
Unfortunately, there is no easy way round the resolution dependence of the number of recombinations 
occurring in the source haloes. It required some experimenting with the evolution of the ionising emissivity
to obtain reasonable reionisation histories for which    hydrogen reionisation is completed at about $z\sim6$  and the evolution of  $\Gamma (z)$   
is close to the observations at redshifts below six. 
As the monochromatic GPU accelerated ATON radiative transfer code  is gratifyingly fast 
(a speedup of 80-100 times compared to CPU-based implementation)
we were able to do this by trial and error.

\begin{figure*}
   \begin{center}
    \begin{tabular}{cc}
      \includegraphics[width=8cm,height=7cm]{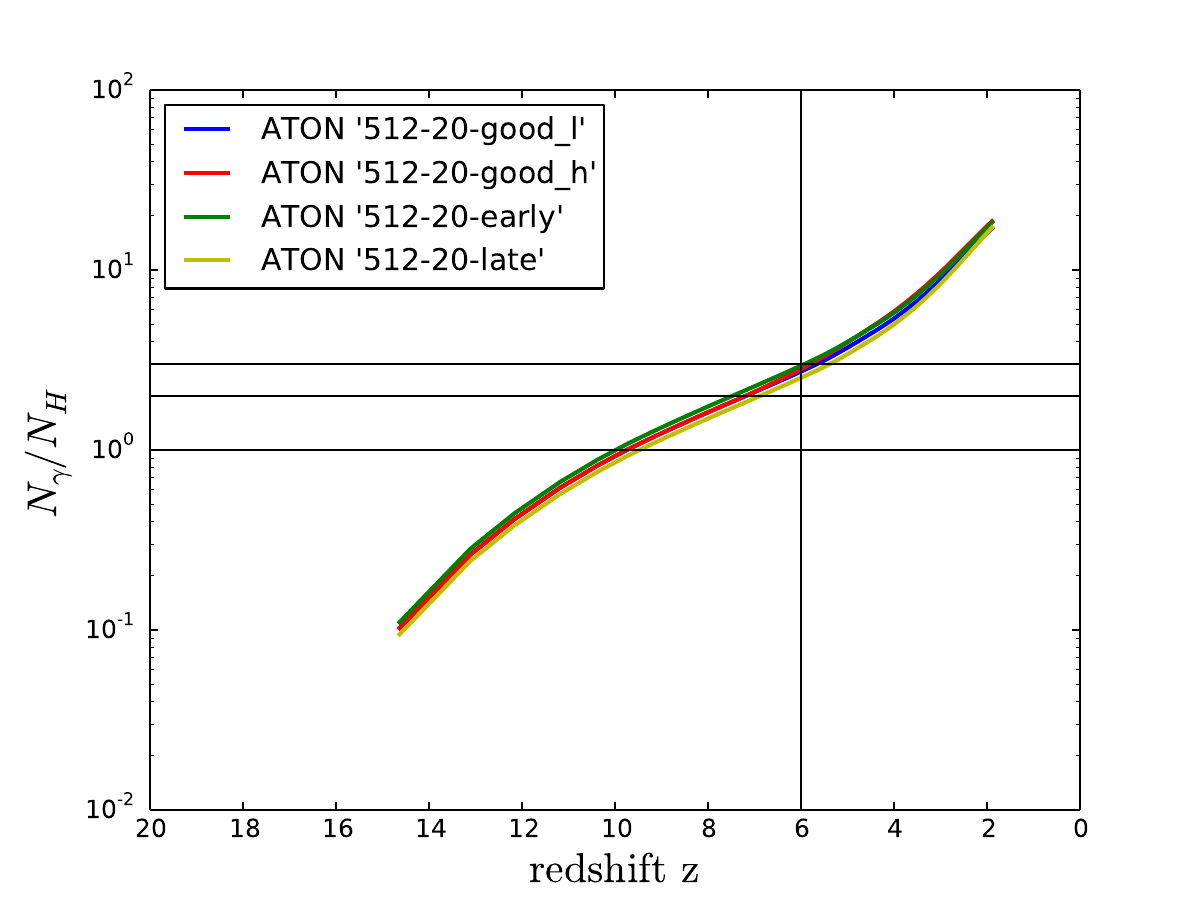} &
      \includegraphics[width=8cm,height=7cm]{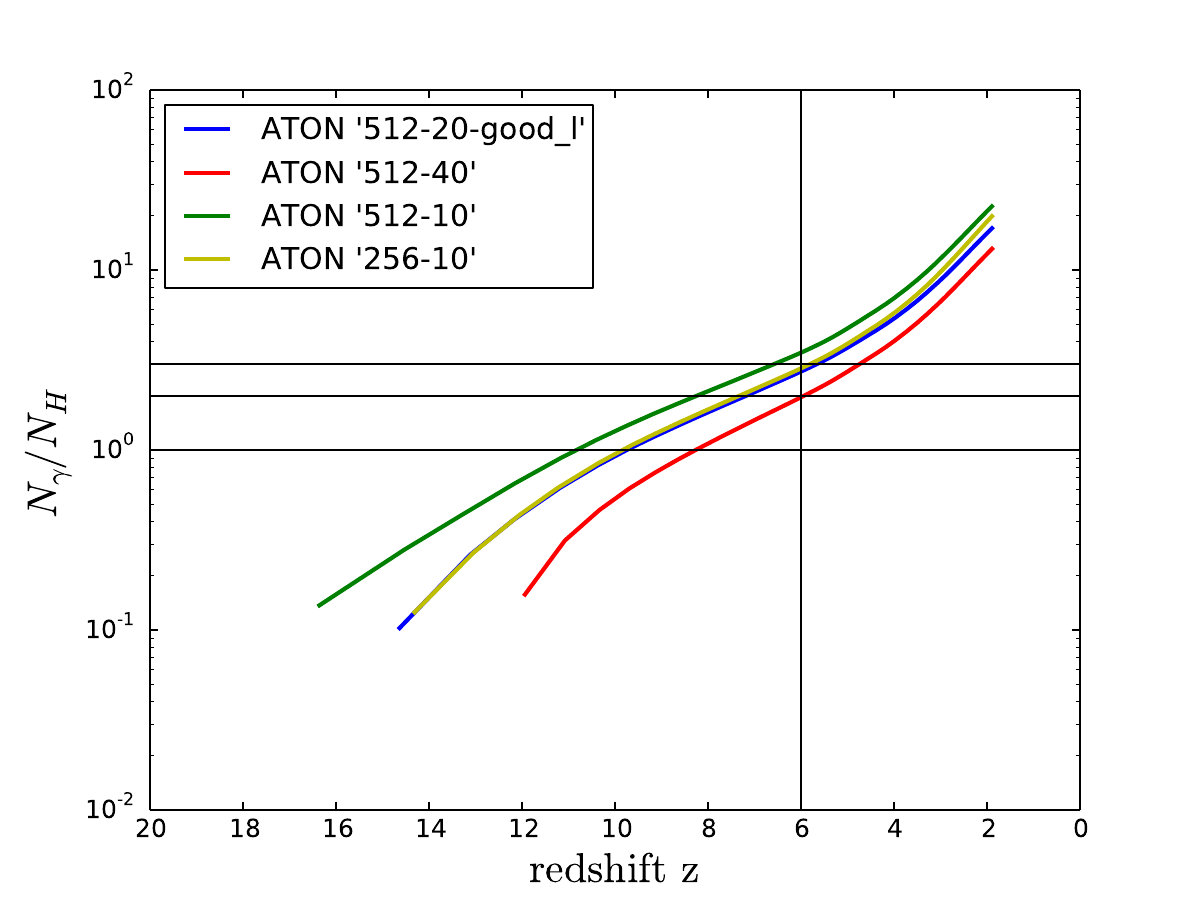} \\
      (a)  & (b)\\
\end{tabular}    
  \caption{(a) The cumulative number of ionising photons emitted per hydrogen atoms as a function of redshift for the ionisation histories in 
  the ATON simulations.
(b): same as (a) but for the simulations with  different box sizes and resolution as described in the text.
The three solid horizontal lines show one, two and three ionising photons emitted per hydrogen atoms at z $\sim$ 6.}
    \label{emissivity_calibration_2}
  \end{center}
 \end{figure*}

\subsection{The simulation set}
\label{Simulation_set}

\renewcommand{\arraystretch}{1.5}
\setlength{\tabcolsep}{0.5cm}
\begin{table*}
\begin{center}
\begin{tabular}{|c||c|c|c|c|c|}
  \hline
  model & 512-20-good\_l & 512-20-good\_h & 512-20-early & 512-20-late \\
  \hline
  resolution  & 512$^3$  & 512$^3$  & 512$^3$  & 512$^3$   \\
  \hline
  box size (Mpc/h) & 20 & 20 & 20 & 20 \\
  \hline
  minimum halo mass ($\mathrm{M_\odot})$ & $1.2\times10^8$ & $1.2\times10^8$ & $1.2\times10^8$ & $1.2\times10^8$ \\
  \hline
    a & 1.3 & 1.3 & 1.4 & 1.2 \\
  \hline
  $z_1$ & 5.8 & 7 & 5.8 & 5.5 \\
  \hline
  $z_2$ & 5 & 5 & 5 & 5 \\
  \hline
  $\alpha_1$ & 0.6 & 0.5 & 0.6 & 0.9 \\
  \hline
  $\alpha_2$ & 1.0 & 1.1 &  1.0 & 0.8\\
  \hline
\end{tabular}
\caption{Basic properties of the simulations for the 512-20 ATON
 simulations with the different reionisation histories.}
\label{tab1}
\end{center}
\end{table*}

\renewcommand{\arraystretch}{1.5}
\setlength{\tabcolsep}{0.5cm}
\begin{table*}
\begin{center}
\begin{tabular}{|c||c|c|c|c|c|}
  \hline
  model & 256-10 & 512-40 & 512-10 \\
  \hline
  resolution  & 256$^3$ & 512$^3$  & 512$^3$  \\
  \hline
  box size (Mpc/h) & 10 & 40 & 10 \\
  \hline
  minimum halo mass  ($\mathrm{M_\odot})$ & $1.9\times10^8$ & $8.9\times10^8$ & $1.6\times10^7$ \\
  \hline
  a & 1.36 & 1.025 & 1.58 \\
  \hline
  $z_1$ & 6 & 5.8 & 6 \\
  \hline
  $z_2$ & 5 & 5 & 5 \\
  \hline
  $\alpha_1$ & 0.75 & 0.5 & 1.0 \\
  \hline
  $\alpha_2$ & 0.85 & 1.0 & 1.0\\
  \hline
\end{tabular}
\caption{Basic properties of the simulations for the simulations with different box sizes and resolution.}
\label{tab2}
\end{center}
\end{table*}

We have run  a suite   of simulations in order to explore the convergence in terms of resolution 
and box size. We also  varied the evolution of the ionising emissivity to study the effect of the timing of the 
overlap of HII regions on the \lya opacity PDF and the spatial fluctuation of the 
photoionisation rate in the post-overlap phase. 

Our standard simulations are based on $512^3$ RAMSES simulations and we have varied the box size of the simulation 
from 10-40 $h^{-1}$Mpc. 
We have also run a $256^3$ simulation to study the effect 
of  resolution and box-size. 

When varying the reionisation history 
we have chosen a default model which reproduces  the measured photoionisation rates 
at $z\sim6$ reasonably well, the `512-20-good\_l' model, where 512 stands for  our default  number 
of RAMSES resolution elements in one dimension and 20  for the 20 $h^{-1}$ Mpc 
default comoving size of the simulation box. 
We have further run simulations where hydrogen is reionised later (model `late') and earlier (model `early')
and a model where overlap occurs at the same redshift as in the `512-20-good\_l' model but with a  somewhat 
larger   ionising emissivity (and thus photoionisation rate) in the phase immediately past overlap (model `good\_h'). 
In tables \ref{tab1} and \ref{tab2} we give the basic parameters of our  RAMSES simulations
as well as  the parameters of  equation \ref{equpowrlaw}.
Fig. \ref{emissivity_calibration_1} shows the redshift evolution of the ionising emissivities used in our simulation suite. 
Fig. \ref{emissivity_calibration_2} shows the corresponding integrated number of ionising photons per hydrogen atom.

\begin{figure}
  \begin{center}
     \includegraphics[width=8cm,height=6.5cm]{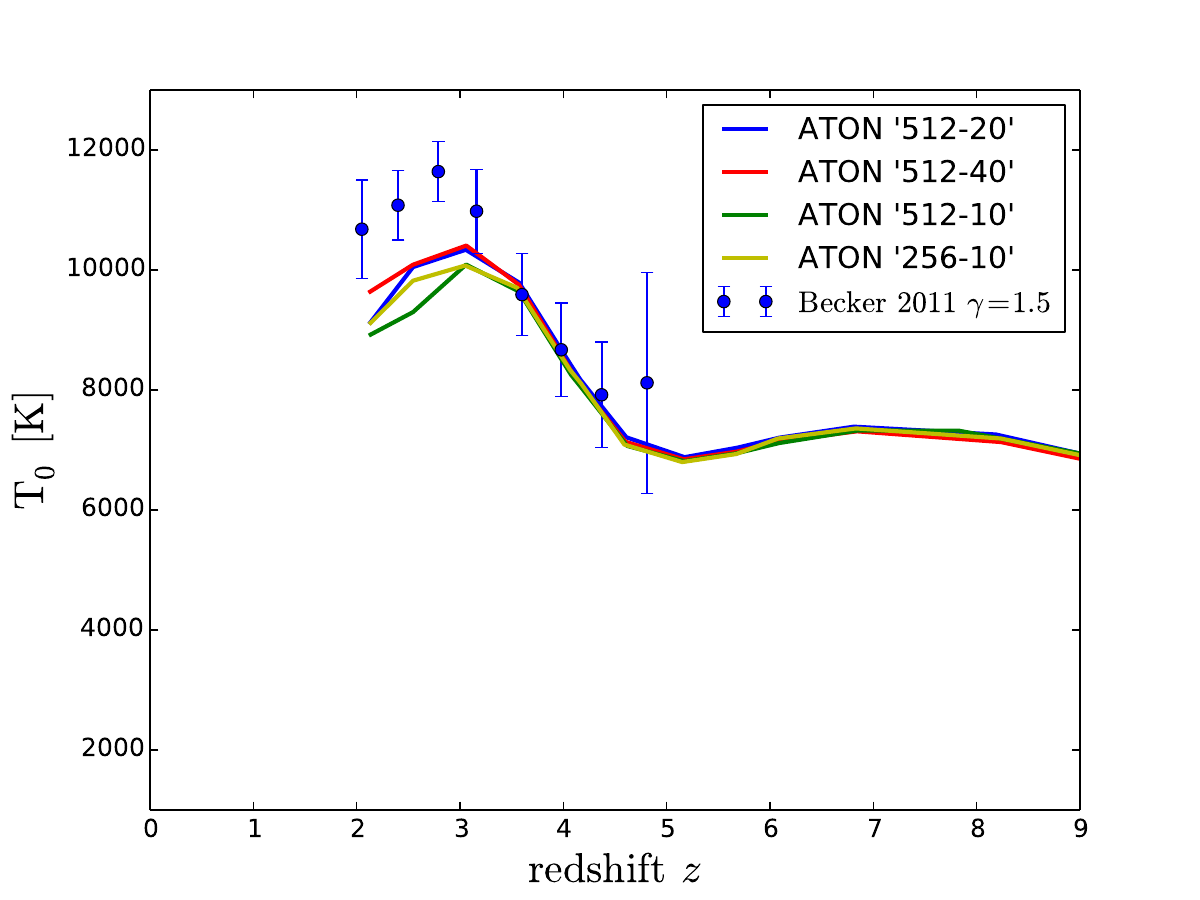} 
  \caption{Evolution of the temperature at mean density $\mathrm{T_0}$ in our different RAMSES/ATON simulations. 
The blue circles with errorbars show  observational constraints from  \citet{2011MNRAS.410.1096B} 
assuming a value of $\gamma = 1.5$ for the slope of the  temperature-density relation. 
The red squares with errorbars show the recent observational value of \citet{2014MNRAS.441.1916B}, again, 
assuming a value of $\gamma = 1.5$ for the slope of the  temperature-density relation.}
    \label{T0_vs_z}
  \end{center}
 \end{figure}

Unfortunately, the temperature calculations as implemented in  ATON simulations   
do not properly  account for the radiative transfer effects on  the temperature as the radiative transfer 
is modelled  monochromatically  and not fully coupled  to the hydro simulation.
As discussed in detail in \citet{2015arXiv1410.1531P},  (optically thin) simulations with 
the HM2012 UV  background model and an equilibrium chemistry solver 
agree actually reasonably well with measured temperatures of the  IGM.
We have therefore assumed the  temperatures obtained in the RAMSES simulations 
for the ATON simulations. 
In Fig.  \ref{T0_vs_z}, we compare the evolution of the temperature at mean density as a function 
of redshift in our  RAMSES/ATON simulations without radiative transfer to measured temperatures from \citet{2011MNRAS.410.1096B} assuming a power law index $\gamma =1.5$ for the temperature-density 
relation of the IGM at the relevant densities.  The agreement is very reasonable.

%

\section{Results}
\label{results}

In this section we describe the main results from our ATON simulations. 
We first focus on the global reionisation history of the different models and 
compare photoionisation rates and neutral hydrogen fractions  to those inferred 
from \lya forest data. 
We then  produce mock  Ly$\alpha$ forest spectra from our simulations 
to compare directly  to observed properties of the \lya forest.
We finally make some predictions of the  UV background fluctuations 
expected in the later stages of the reionisation from our simulations.

\subsection{Reionisation history}

\subsubsection{Evolution of the  neutral hydrogen fraction}

\begin{figure*}
   \begin{center}
      \includegraphics[width=\textwidth,height=\textheight,keepaspectratio]{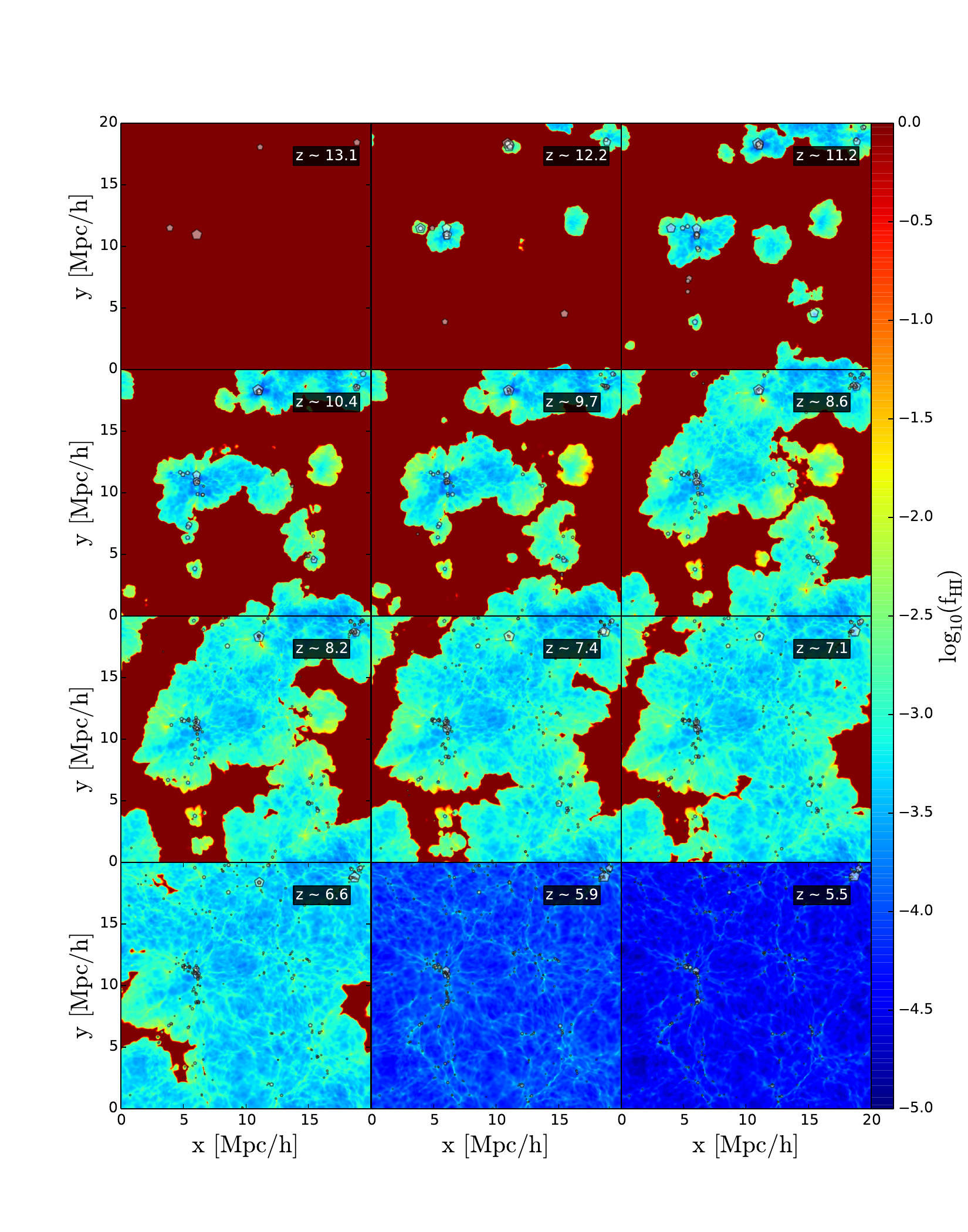}    
  \caption{Evolution of the spatial distribution of the neutral fraction for the `512-20-good\_l' model for a slice of 39.06 comoving kpc thickness in the mid plane of the simulation.  
The white pentagons indicate the location of the dark matter haloes assumed as ionising sources with the size scaled to the mass of the halo.
}
    \label{red_seq_512_10_good}
  \end{center}
 \end{figure*}

Figure \ref{red_seq_512_10_good} shows the  evolution of the spatial distribution of the neutral fraction for the `512-20-good\_l' model.
Individual HII regions start to appear as early as $z\sim16$ and by $z\sim10$  
there is considerable overlap of HII regions. Contiguous regions  of   still neutral gas in the post-overlap phase 
persist to about  $z\sim7$. Later on fully neutral gas is expected to be mostly confined to virialized haloes, 
however the ionising sources in our simulations appear to keep the gas in their host haloes fully ionised. 
Note that this  may  be an artefact as the simulation still don't  have the resolution to properly resolve the ISM 
in these haloes. We also do not include dust and assume that the haloes continuously emit ionising radiation.
Note also that we do not have the resolution to properly resolve the majority of mini-haloes with virial temperatures 
below the atomic cooling threshold. The smallest haloes in our 512-20 simulation 
have circular velocities of  $v_{\mathrm{circ}}\sim10.7$ km/s somewhat above the atomic cooling threshold, 
but even at this resolution  the simulations are already  significantly  incomplete for these haloes
(see Appendix~\ref{halo_mass_func} for halo mass functions for our simulation suite).  
Finally note again that the effect of the reionisation heating is only taken into account 
by the equilibrium modelling with the HM2012  UV background model in the RAMSES simulations 
without  radiative transfer.

\begin{figure*}
  \vskip -1.0truecm
   \begin{center}
      \includegraphics[width=\textwidth,height=\textheight,keepaspectratio]{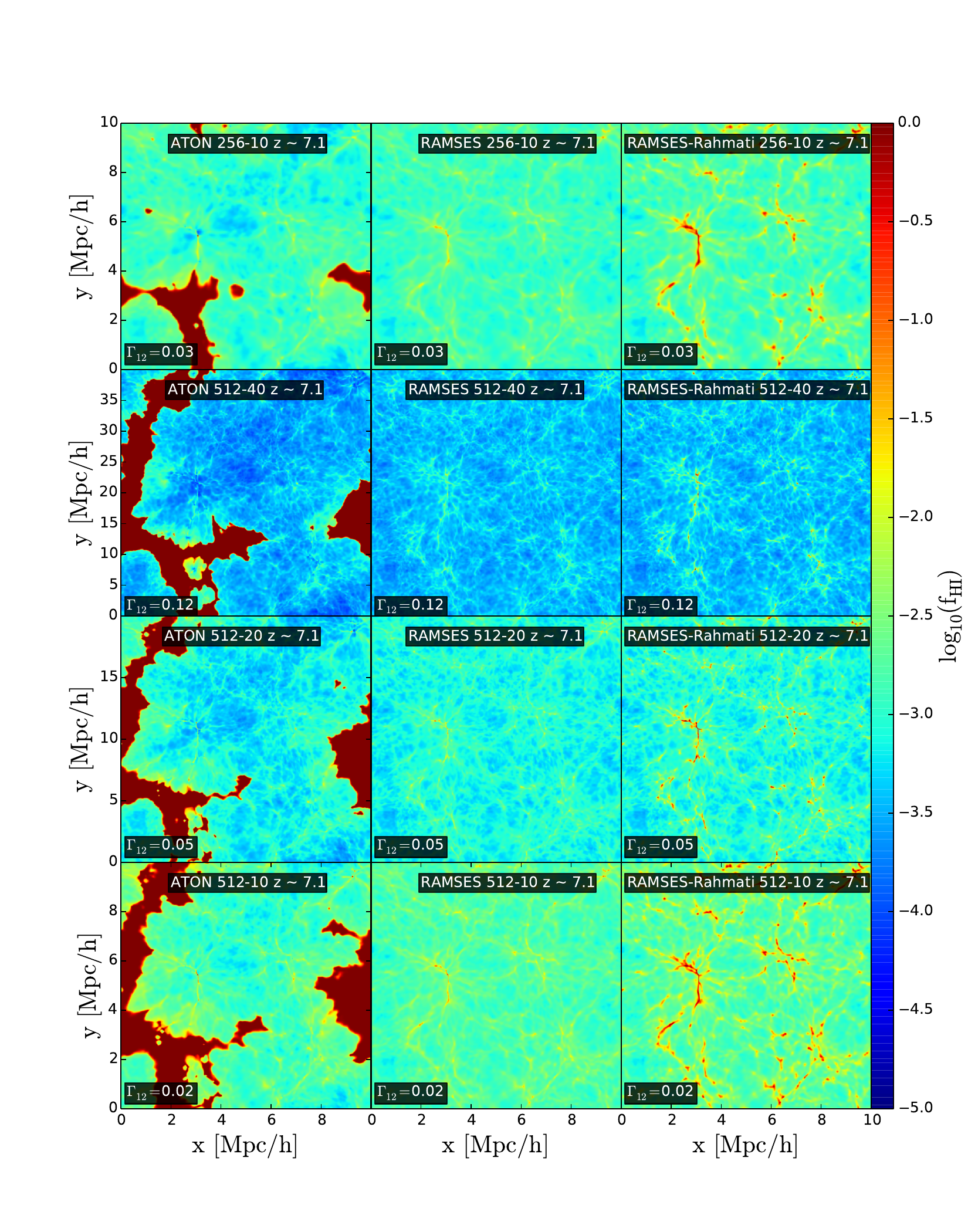}
  \caption{Spatial distribution of the neutral fraction at $z = 7.1$ : First, second, third and fourth rows are for the `256-10' simulation, 
the `512-40' simulation, the `512-20-good\_l' simulation and `512-10' simulation, respectively.
The first column  shows the full radiative transfer simulations with ATON.  
The second column  shows the optically thin  RAMSES simulations 
with photoionisation rate rescaled to match that of the corresponding  ATON  simulation in the first column. The third column shows the RAMSES outputs 
post-processed with the model/fit of \citet{2013MNRAS.430.2427R} to account for self-shielding to ionising photons. 
The photoionisation rate in the slices shown is indicated on the plot.}
    \label{map_neutral_z7}
  \end{center}
 \end{figure*}

In  Fig. \ref{map_neutral_z7} we compare the spatial distribution of ionised and neutral regions at
$z\sim7.1$ of our ATON  simulations (left column)  for a range of resolutions and box sizes.
The ionisation maps are resolution dependent and in particular the remaining neutral gas in the haloes in 
already ionised regions  is  increasing with increasing resolution as the gas in the individual dark matter haloes becomes 
better resolved and is able to recombine and self-shield due to higher densities.
We show  the corresponding RAMSES simulations without radiative transfer in the  middle column. 
To make this a fair comparison we have rescaled the photoionisation rate in the RAMSES 
simulation to be the same as the mean photoionisation rate in the ionised regions 
in the slice of the ATON simulation shown in the left column.  

We have  furthermore  implemented the self-shielding  correction suggested by
\citet{2013MNRAS.430.2427R} 
based on their radiative transfer simulations  in the RAMSES simulations without radiative transfer (shown in the right column).
In the \citet{2013MNRAS.430.2427R} prescription the self-shielding of the gas is taken into account by  a density-dependent 
photoionisation rate, obtained by an empirical fit to  their radiative transfer simulations (\citealt{2013MNRAS.430.2427R}),

\begin{equation}
\begin{split}
 \frac{\Gamma_{\mathrm{ss}}}{\Gamma}&=0.98\times\left[1+\left(\frac{\Delta}{\Delta_{\mathrm{ss}}}\right)^{1.64}\right]^{-2.28}  \\
 & + 0.02\times\left[1+\frac{\Delta}{\Delta_{\mathrm{ss}}}\right]^{-0.84}
\end{split}
\end{equation}

where $\Gamma_{\mathrm{ss}}$ is the value of the photoionisation rate in a cell of the computational box after accounting for self-shielding while $\Gamma$ 
is the value of the background photoionisation rate before accounting for self-shielding.
$\Delta_{\mathrm{ss}}$ is the overdensity above which the gas begins to self-
shield (see \citealt{2013MNRAS.429.1695B}),

\begin{equation}
 \Delta_{\mathrm{ss}}=36\times\Gamma_{12}^{2/3}T_4^{2/15}\left(\frac{\mu}{0.61}\right)^{1/3}\left(\frac{f_e}{1.08}\right)^{-2/3}\left(\frac{1+z}{8}\right)^{-3}
\end{equation}

where, $\Delta = \rho / \overline{\rho}$ is the overdensity, $\Gamma_{12} = \Gamma/10^{-12} \mathrm{s^{-1}}$ is the
background photoionisation rate, $T_4 = T /10^4 K$, with
$T$ the temperature of the gas, $\mu$
is the mean molecular weight and $f_e=n_e/n_H$
is the free electron fraction with respect to hydrogen.

We have chosen $z=7.1$ for this comparison as there is particular interest in this redshift 
due to the highest redshift QSO ULAS J1120+0641 at $z=7.085$  which may show signs of a red damping wing 
in its absorption spectra (\citealt{2011Natur.474..616M}, \citealt{2011MNRAS.416L..70B}, but see Bosman et al. 2015 (in prep)). 
Note that while there are still substantial contiguous neutral regions 
extending well beyond the extent of individual haloes  in the ATON simulations the volume filling 
factor of the neutral gas in the optically thin simulations is considerably smaller and is 
confined to the haloes even when self-shielding with the \citet{2013MNRAS.430.2427R} prescription is taken into 
account. The self-shielding of dense gas in haloes located in already ionised regions of the ATON simulations is 
considerably weaker than predicted by the \citet{2013MNRAS.430.2427R} model which does not account for the local ionising 
sources located in these haloes which are emitting with constant emissivity in our models.

\begin{figure*}
   \begin{center}
      \includegraphics[width=\textwidth,height=\textheight,keepaspectratio]{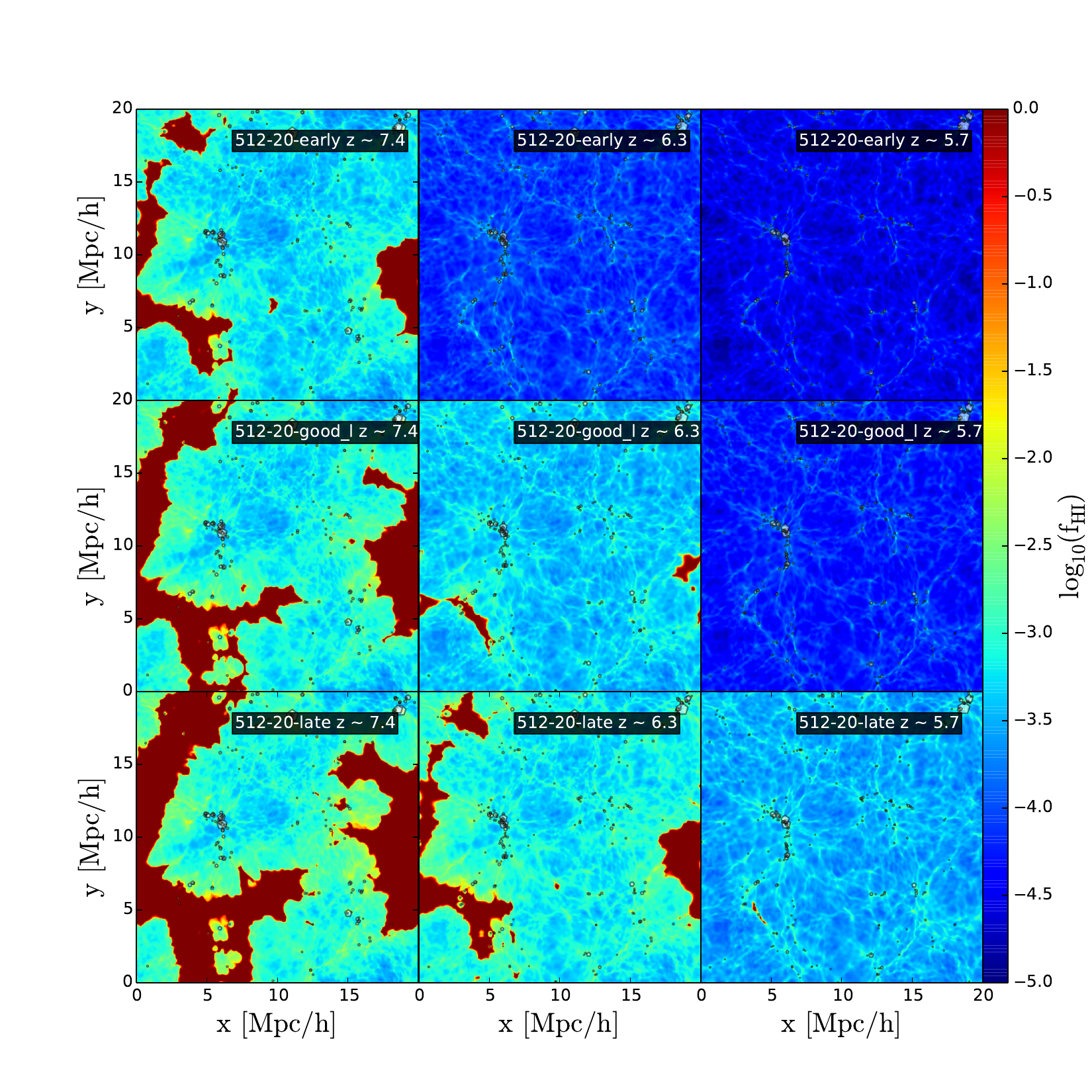}    
  \caption{The spatial distribution of the neutral fraction at $z=(7.5,6.3,5.7)$ for the different ionisation histories 
  in our ATON simulations.  First, second and third rows corresponds to the `512-20-early', `512-20-good\_l' and the `512-20-late' models.
The white pentagons indicates the location of the dark matter haloes assumed as ionising sources with the size scaled to the mass of the halo. 
}
    \label{map_neutral_10_mpc}
  \end{center}
 \end{figure*}

\begin{figure*}
   \begin{center}
    \begin{tabular}{cc}
      \includegraphics[width=8cm,height=7cm]{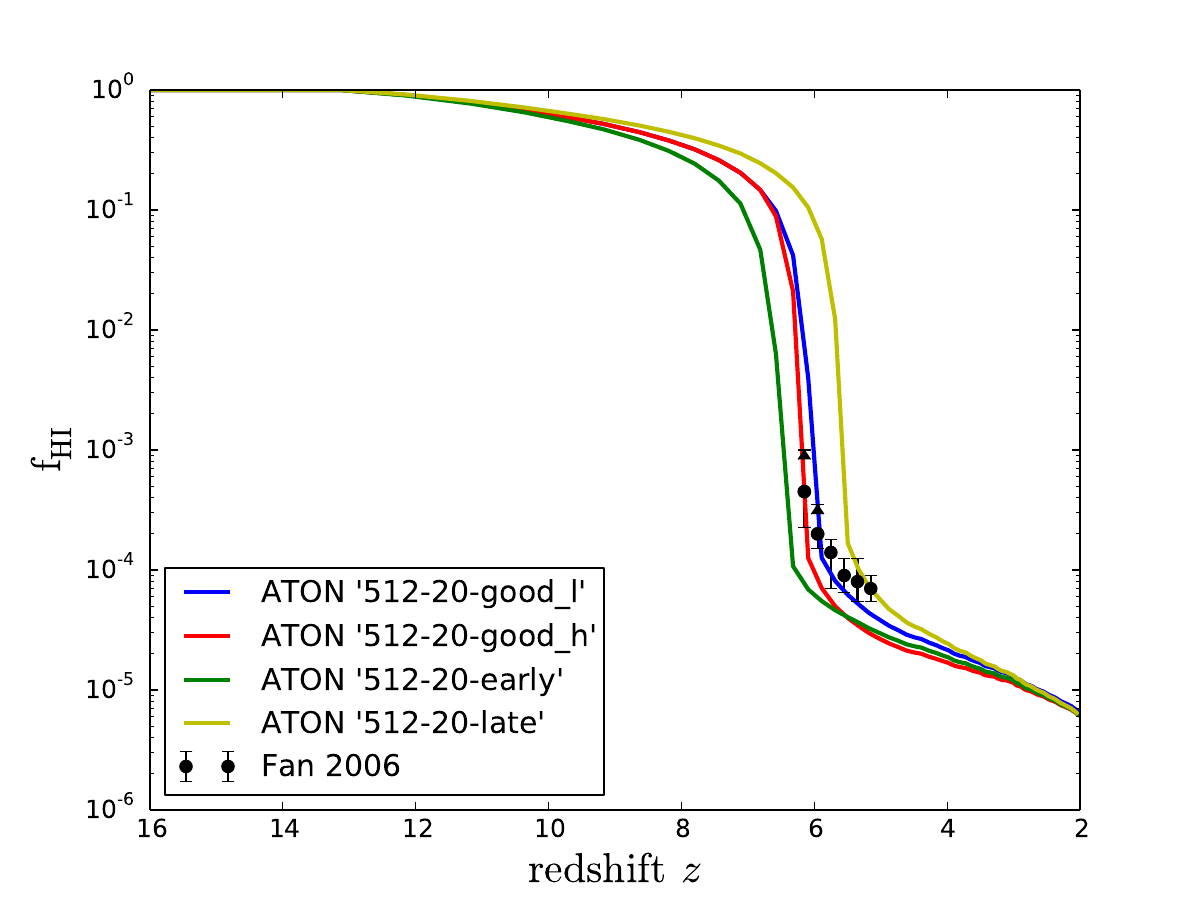} &
      \includegraphics[width=8cm,height=7cm]{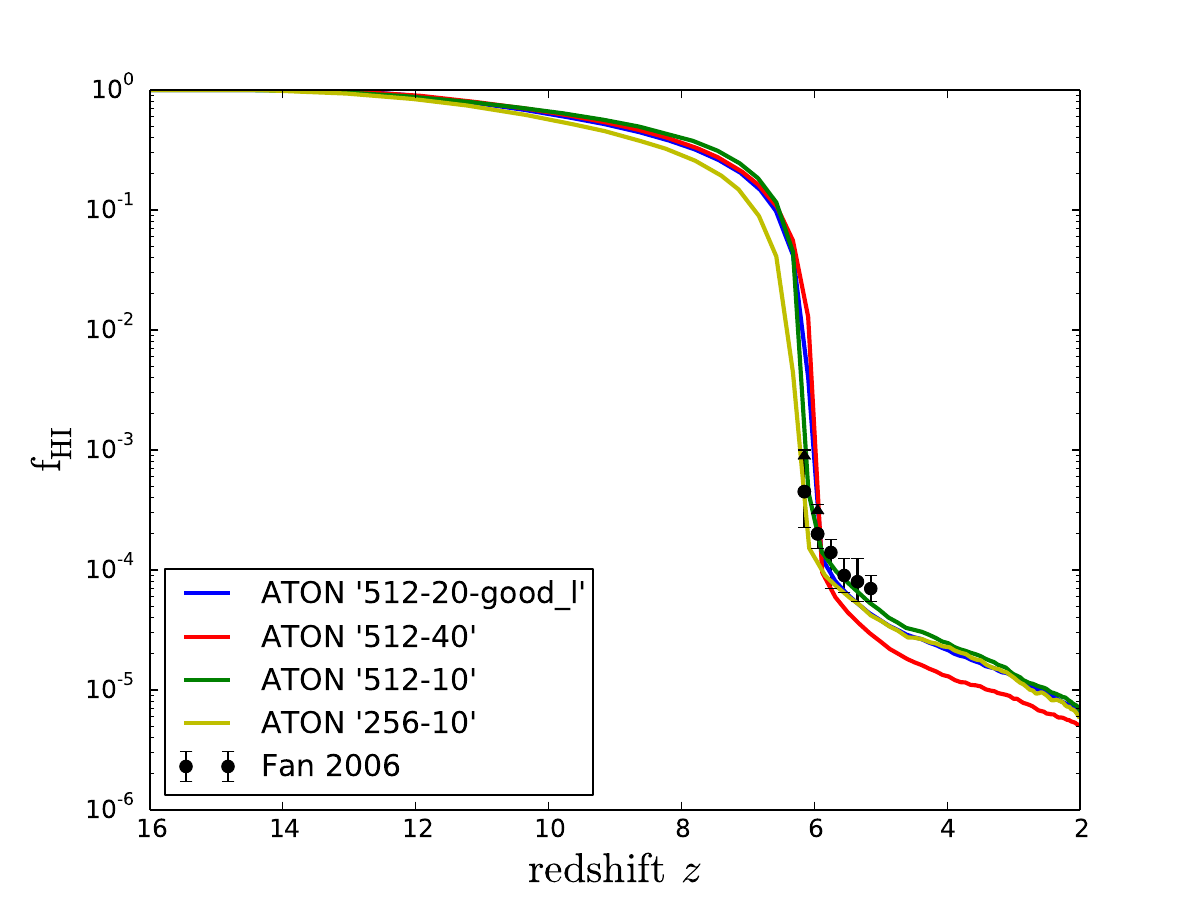} \\
      (a)  & (b)\\
      \includegraphics[width=8cm,height=7cm]{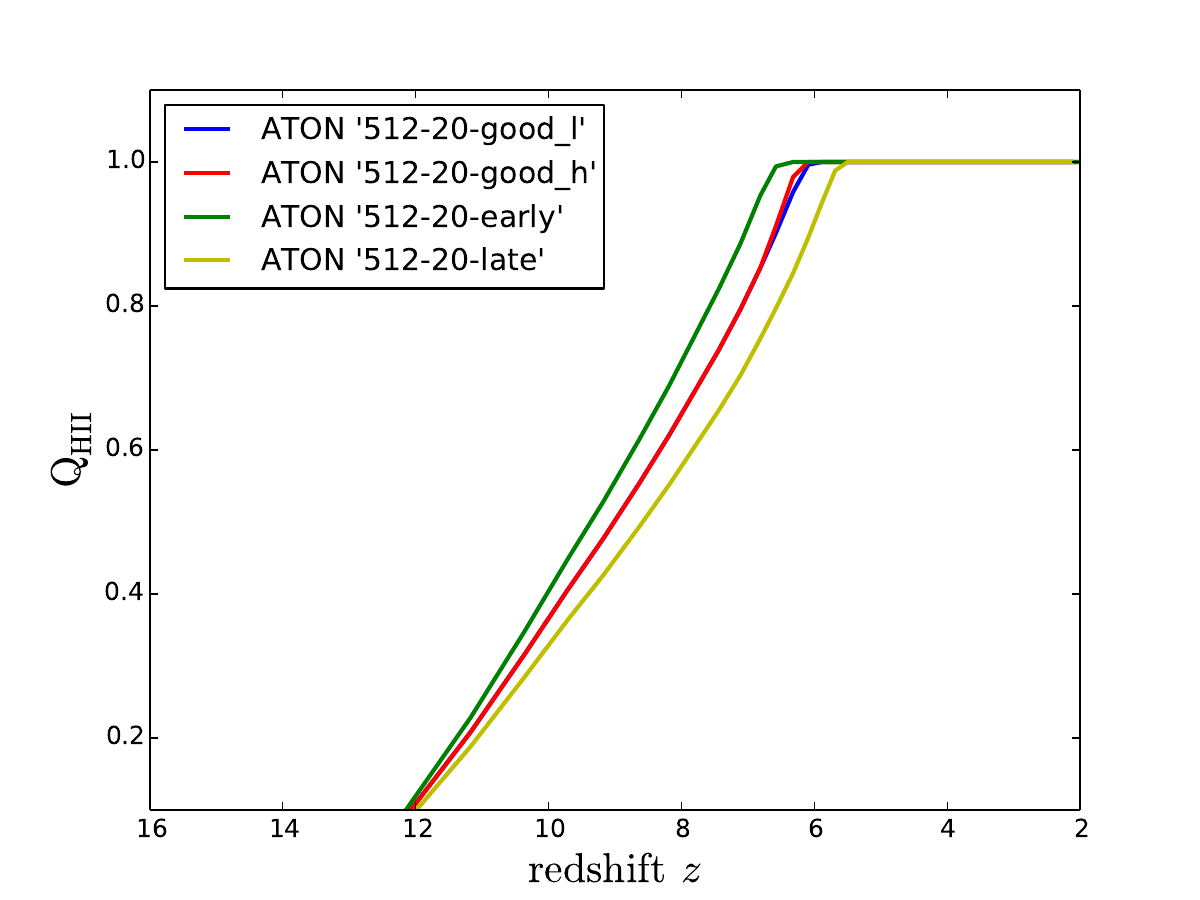} &
      \includegraphics[width=8cm,height=7cm]{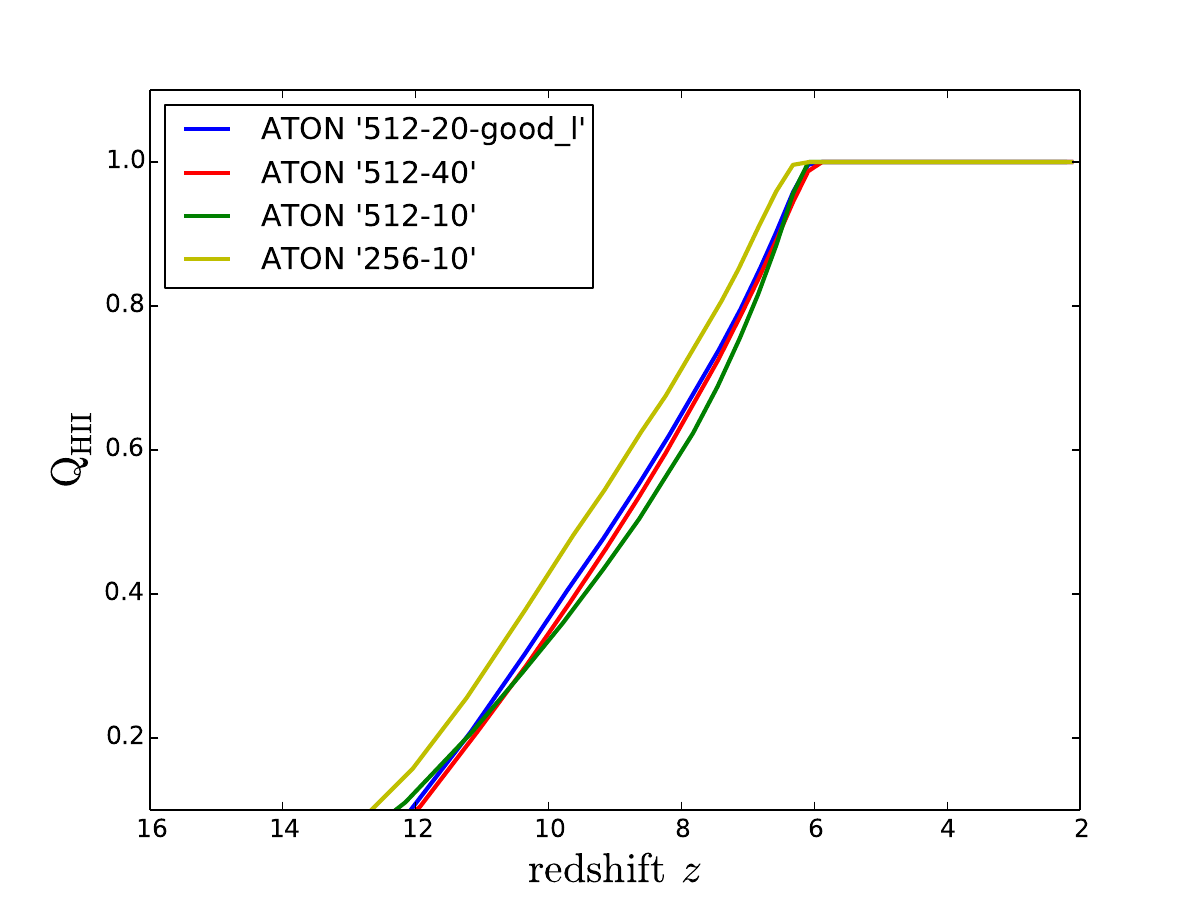} \\
      (c)  & (d)\\
\end{tabular}    
  \caption{(a) Evolution of the mean volume-weighted neutral fraction for the  ATON simulations with different reionisation histories,
(b) same as (a) for the simulations with different box sizes and resolutions,
(c) evolution of the volume filling factor of ionised regions (with an ionisation fraction $\ge0.5$) for the  ATON simulations with different reionisation histories,
(d) same as (c) for the simulations with different box sizes and resolutions. The data points are from \citet{2006AJ....132..117F} 
and are based on an analytical model  for the opacity  PDF.
   }
    \label{x_vs_z}
  \end{center}
 \end{figure*}

Figure \ref{map_neutral_10_mpc} shows the evolution of the spatial distribution of the neutral fraction  at three redshifts for the four  
`512-20' models with the corresponding emissivities 
reported in Table~\ref{tab1}. As expected, increasing/decreasing  the emissivities 
accelerates/delays reionisation.
The different models bracket a reasonable  range of reionisation redshifts.
The differences in the timing of the reionisation process between the different models are further  quantified 
in the upper left panel of Figure~\ref{x_vs_z} that shows the evolution of the mean neutral fraction, while the bottom left panel of the 
same figure shows the evolution of the volume filling factor for our suite of simulations.
As already mentioned our `good' models  
were chosen such that hydrogen reionisation is mostly  completed by $z=6-6.5$. At  this redshift 
the still neutral regions cease  to hang together and become separate islands
and the mean neutral fraction drops rapidly by several orders of magnitudes to 
a level of $f_{\rm HI} \approx 10^{-4}$, while the volume filling factor of ionised regions rises to $\mathrm{Q_{HII}}=1$.  
The same happens at somewhat lower/higher  redshift in our late/early model.

\begin{figure*}
   \begin{center}
    \begin{tabular}{cc}
      \includegraphics[width=8cm,height=7cm]{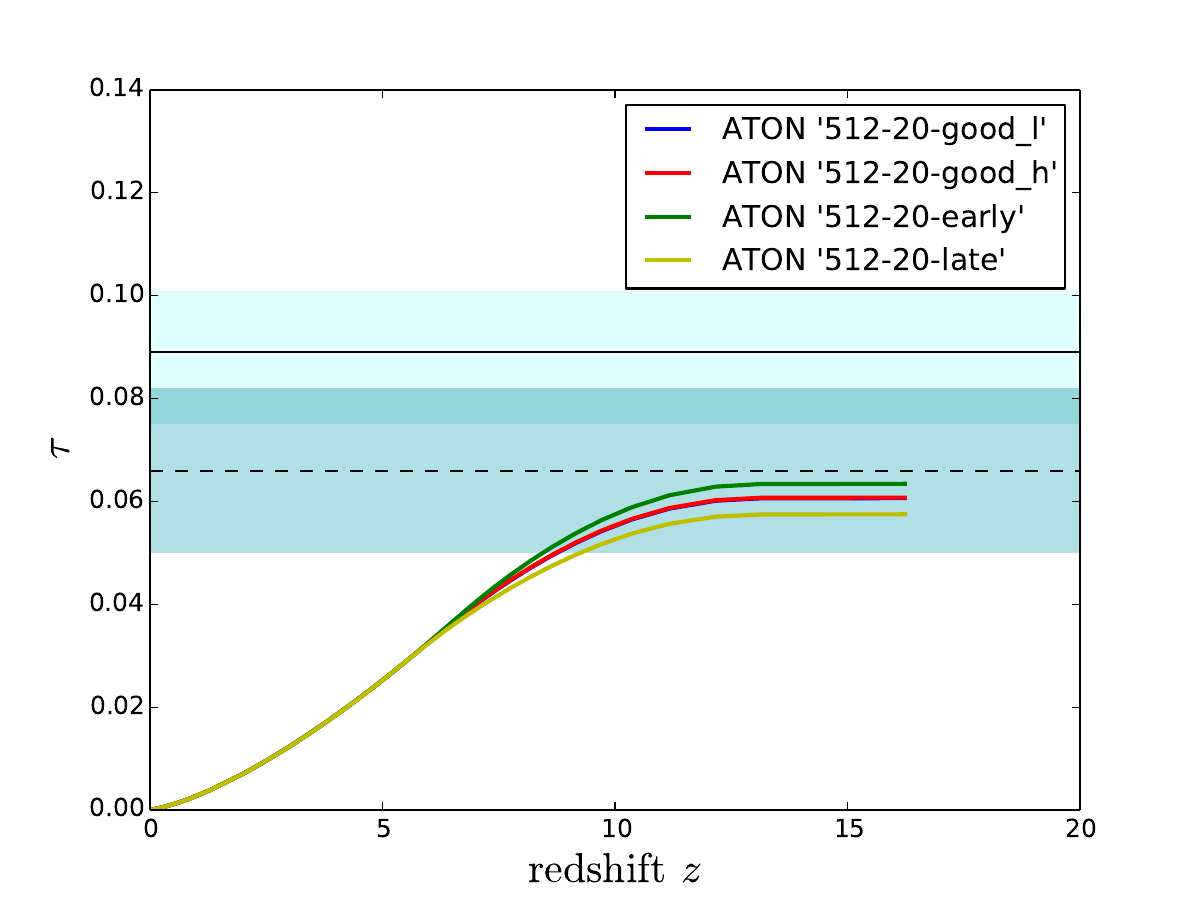} &
     \includegraphics[width=8cm,height=7cm]{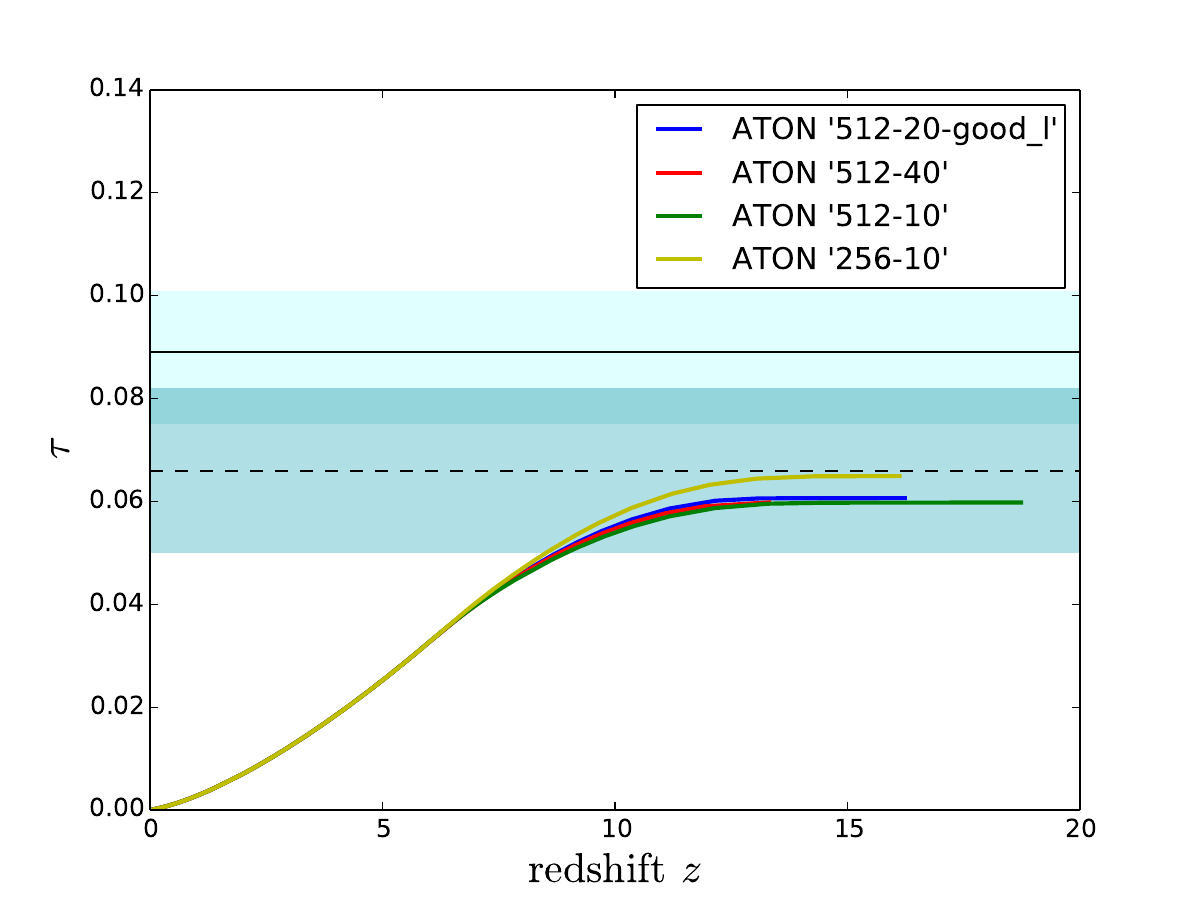}  \\
     (a) & (b)\\
\end{tabular}    
  \caption{(a) The integrated Thompson optical depth for the  ATON
 simulations with the different reionisation histories.
(b): Same as (a) for the simulations with different box sizes and resolution.
The solid black line with the light blue area represents the 68\% confidence limits
from the 2013 Planck+WMAP data (see \citealt{2014A&A...571A..16P}).          
The dashed black line with the shaded dark blue area represents the 68\% confidence limits
from the Planck 2015 TT+lowP+lensing data release (see \citealt{2015arXiv150201589P}).
}
    \label{tau_thompson}
  \end{center}
 \end{figure*}

Figure \ref{tau_thompson} shows the integrated  Thompson optical depth  
for our  simulation suite,

\begin{equation}
\tau(z) = c\sigma_t \int_z^0 n_e(z)\frac{dt}{dz}dz,
\end{equation}

where $\sigma_t$ is the Thompson cross section of the electron, and
$n_e(z) $  is the electron density.
At low redshift, where the Universe is fully ionised 
the curves all converge. Our reionisation histories are all  somewhat  below the mean value of the 2013 Planck+WMAP data 
results and are in excellent  agreement with the Planck 2015 TT+lowP+lensing results.
There is little dependence of the  Thomson optical depth on resolution as we have adjusted the emissivity 
to obtain similar reionisation redshifts. The slightly earlier reionisation in the `256-10' model results 
in a somewhat larger optical depth as expected.  
Note that we have assumed here that He becomes singly ionised in regions where hydrogen 
is ionised. Note further that the reionisation history at high redshift (say $z>10$) is poorly constrained 
by our comparison to \lya forest data. There should thus exist models with a different evolution of the ionising emissivity at very high redshift 
(where we have not modified here the redshift evolution  assumed in HM2012)
that are equally consistent with the \lya forest data and have a different Thomson optical depth.

\subsubsection{Photoionisation rate and mean free path of ionising photons}
\label{gamma_mfp_vs_z}

\begin{figure*}
   \begin{center}
    \begin{tabular}{cc}
      \includegraphics[width=8cm,height=7cm]{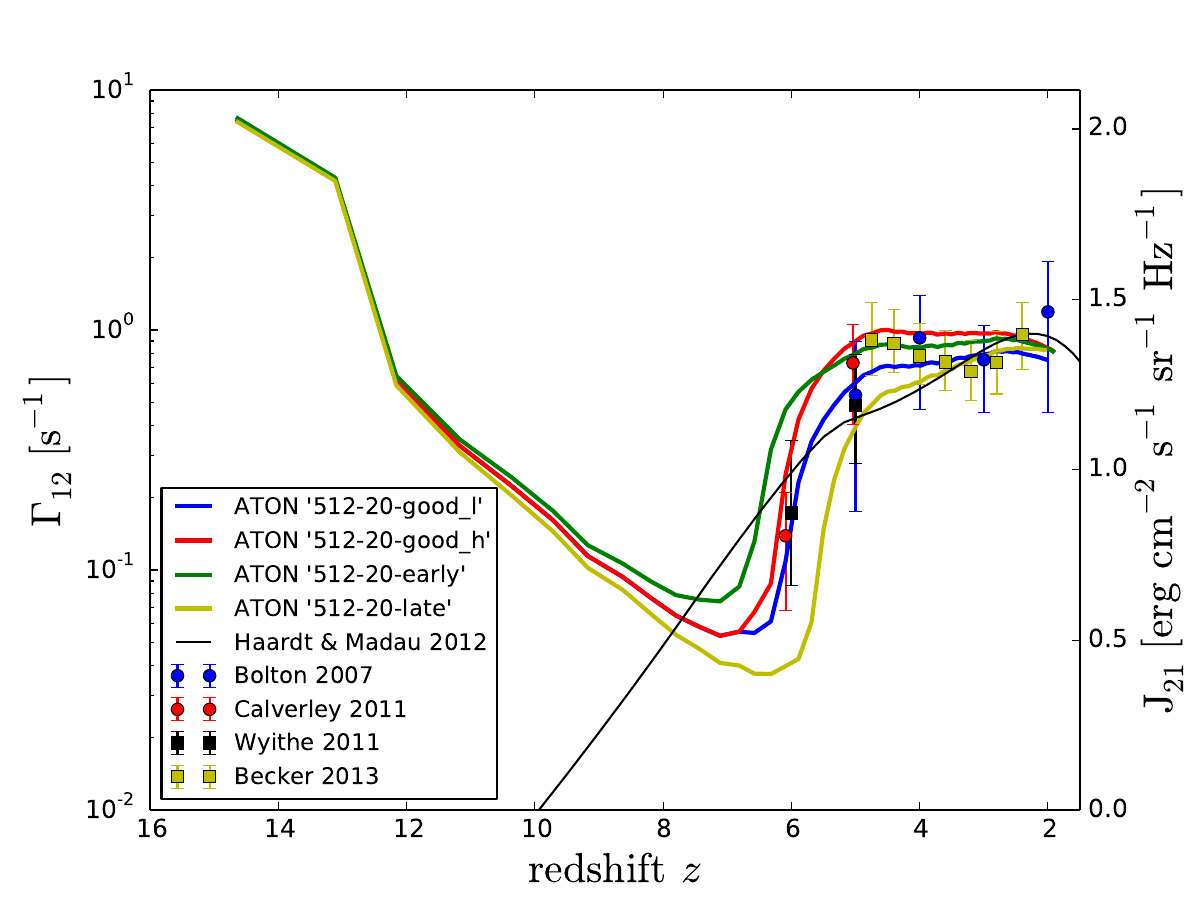} &
      \includegraphics[width=8cm,height=7cm]{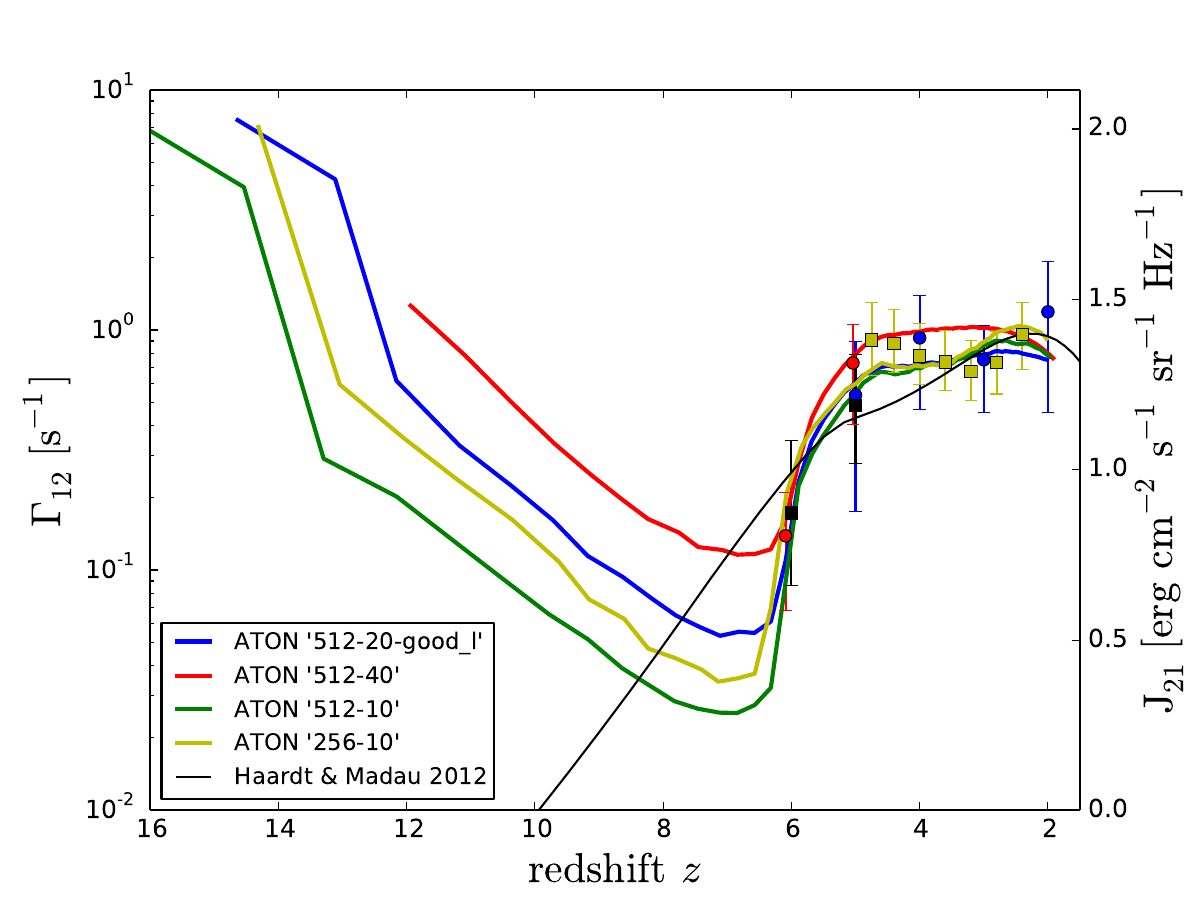} \\
      (a)  & (b)\\
\end{tabular}    
  \caption{ (a) Evolution of the hydrogen photoionisation rate for the ATON simulations with different reionisation histories 
(computed in ionised regions with an ionisation fraction $\ge0.5$).
(b) same as (a) for the simulations with different box sizes and resolutions.
The solid black line shows the evolution of the photoionisation rate in the  uniform HM2012 UV background  model. 
The observational constraints are from \citet{2007MNRAS.382..325B}, \citet{2011MNRAS.412.2543C}, \citet{2011MNRAS.412.1926W}  
and \citet{2013MNRAS.436.1023B}. Note that the data points have been rescaled (by factors ranging from 0.85 to 1.03) to 
that inferred for the  cosmological parameters  and temperature-density relation in our RAMSES simulations 
with the scaling relation in \citet{2005MNRAS.357.1178B} and \citet{2007MNRAS.382..325B}. }

    \label{tau_j21_vs_z}
  \end{center}
 \end{figure*}

\begin{figure*}
   \begin{center}
    \begin{tabular}{cc}
      \includegraphics[width=8cm,height=7cm]{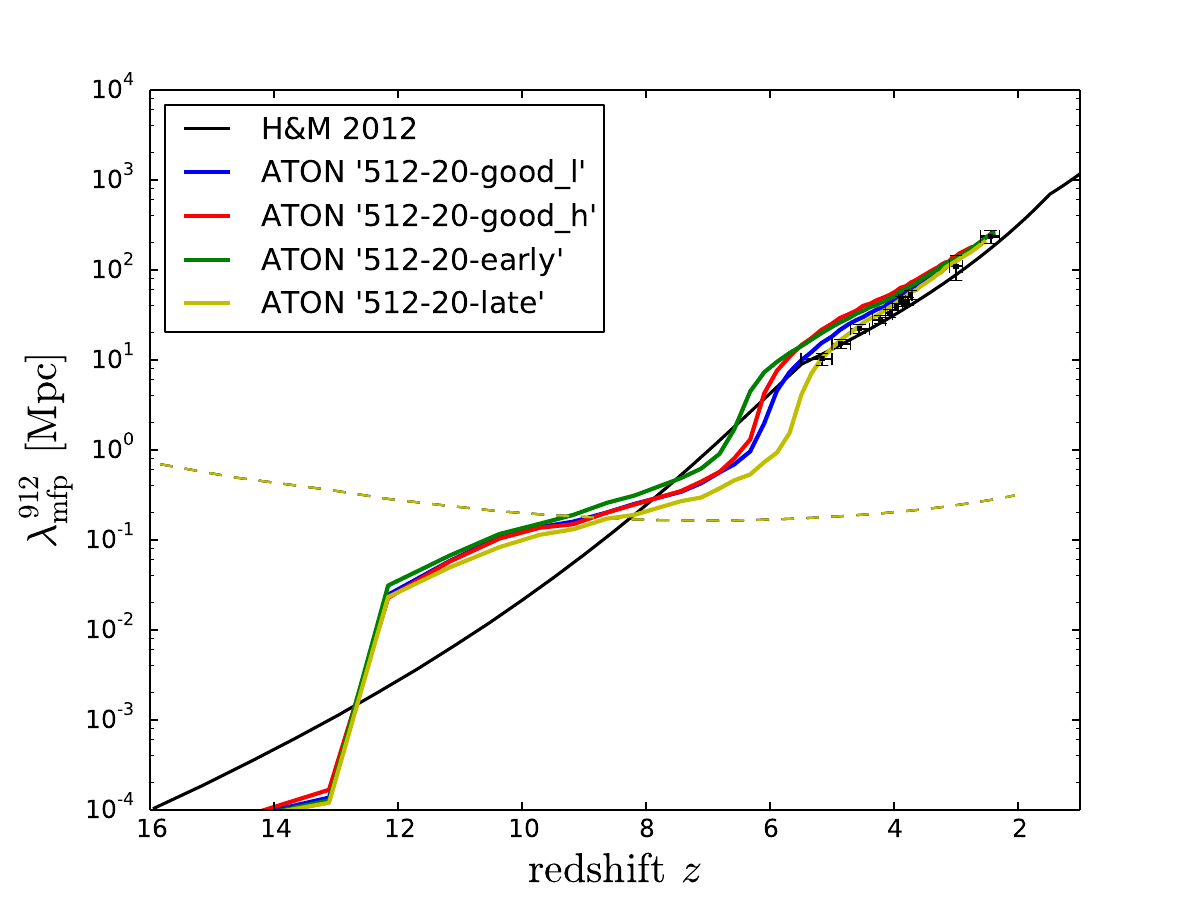} &
      \includegraphics[width=8cm,height=7cm]{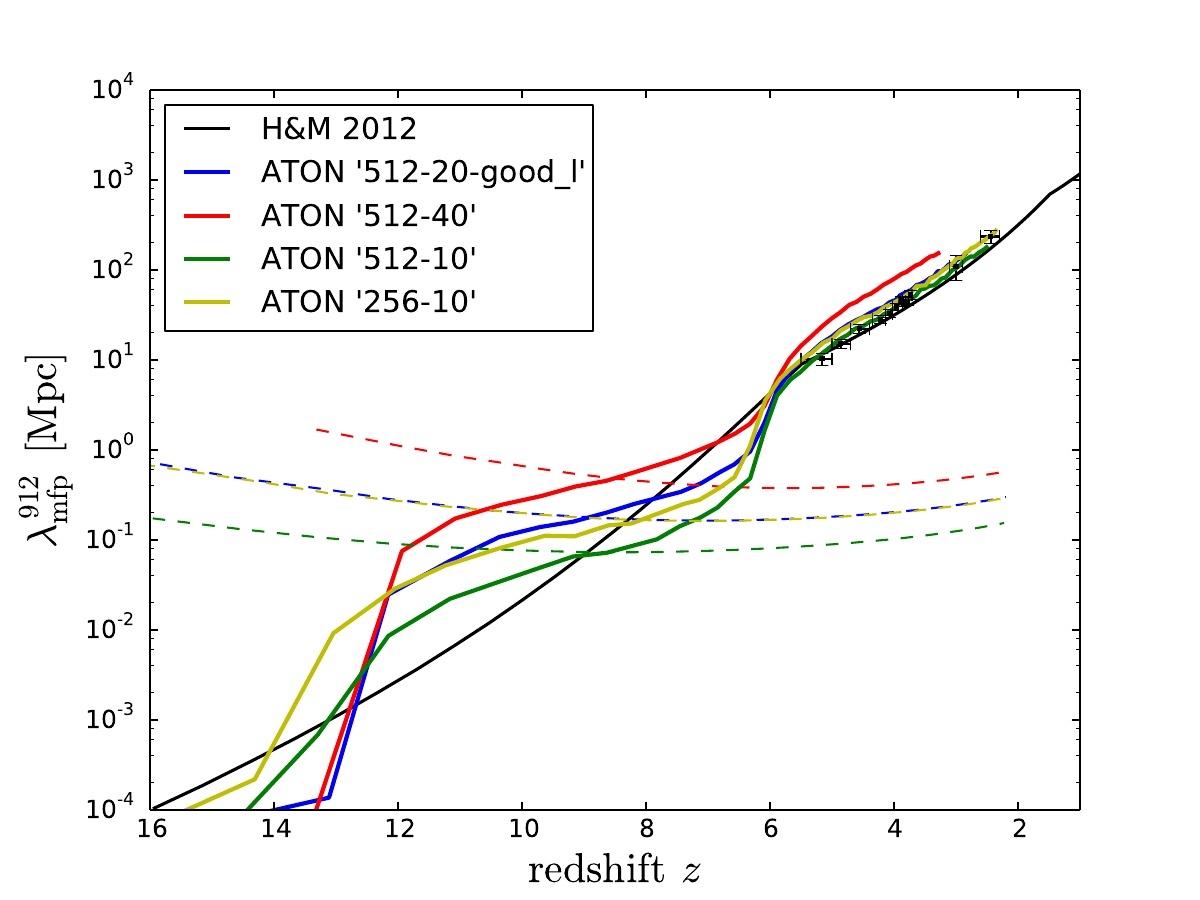} \\
      (a)  & (b)\\
\end{tabular}    
  \caption{
(a) Evolution of the space averaged mean free path of Lyman limit ($912 \, \mathrm{\AA}$) photons (in proper Mpc) 
for the ATON simulations with different reionisation histories (including both ionised and neutral regions). 
(b) same as (a) for the simulations with different box sizes and resolutions.
The  dashed curves  show the evolution of  the mean distance between ionising sources  
in the different ATON  simulations.
The black squares with errorbars show observational constraints from \citet{2013ApJ...765..137O} at z = 2.44, \citet{2013ApJ...775...78F} at
z = 3.00, \citet{2009ApJ...705L.113P} at z = 3.73, and
\citet{2014MNRAS.445.1745W} at z = 4.56 as compiled by
\citet{2014MNRAS.445.1745W}. The black solid line shows the evolution in the HM2012 model.}
    \label{mfp_vs_z}
  \end{center}
 \end{figure*}

The coloured curves in Fig. \ref{tau_j21_vs_z} show the  evolution of the space averaged hydrogen photoionisation rate for  
our ATON simulations.
Note that we have excluded not yet ionised regions when calculating the average photoionisation rates.
The photoionisation rate is first decreasing as the HII regions expand and the distances from the ionising sources 
increases and it then increases once overlap is achieved and multiple sources contribute to the local photoionisation rate.
By moderately scaling our emissivities with regard to  the HM2012 UV background model we  succeeded in  reproducing the  
observed photoionisation rates rather well.  
The only exception is the `512-20-late' model that has photoionisation rates   at $z\ge5$ somewhat lower than observed. 
Reionisation appears indeed to happen too late in this model to be consistent with the  observed photoionisation rates as we had intended for this model. 
Tuning  the ionising emissivities to reproduce the observed apparent  lack of evolution of the photoionisation rate at $6>z>2$ in the other models 
required  a significantly shallower rise of the ionising emissivity compared to  the HM2012 model at $5>z>2$ (see Fig. 1) that assumes a rather rapid 
rise of the ionising emissivity due to QSOs in this redshift range (see Tables \ref{tab1} and \ref{tab2} for the coefficients of the power law of equation \ref{equpowrlaw}). 
Note that  \citet{2014arXiv1410.2249M}  came to similar conclusions  based on their analytic modelling of  photoionisation rates. 
Our  `good\_l' and `good\_h' models nicely bracket  the inferred  photoionisation rate   at $6>z>5$ where the photoionisation rate  is     
most difficult to measure and also evolves rather  rapidly.

To better understand the evolution of the photoionisation rates it is illustrative to also have a look at the evolution of the mean free path of ionising photons
relative to the mean distance between its sources which we show in  Fig.  \ref{mfp_vs_z}. In all models the mean free path 
shown by the solid  coloured curves is initially  shorter than the mean distance between ionising sources and rises slowly with decreasing  redshift 
until mean free path and mean distance between sources become similar. Soon afterwards  the ionised regions fully percolate 
and the mean free path rises very rapidly.
The evolution of the photoionisation  rate then scales as the product of ionising emissivity 
and mean free path, $\Gamma \propto \epsilon_{\rm ion} \lambda_{\rm mfp}$. 
The photoionisation rate  therefore shows a similar rapid rise when 
the ionised regions fully percolate. This behaviour of $\Gamma$ and $\lambda_{\rm mfp}$ has already been discussed 
for the  early  cosmological radiative transfer  simulations by \citet{2000ApJ...535..530G}. 
There is good agreement with observations of the mean path as compiled 
by \citet{2014MNRAS.445.1745W}, but note that as discussed by \citet{2013MNRAS.436.1023B} at $z\la4$  
the approximation  made here that the mean-free path is small compared to the Hubble radius becomes 
rather poor. At $z\la 4$ redshifting of  ionising photons between emission and absorption has  to be taken 
into  account for a proper comparison of the  mean free path predicted by the simulation and that inferred from \lya forest data.

For reference we also show the mean free path 
corresponding to the assumed opacity in the HM2012 UV background model. 
There is a clear trend of increasing mean free path (as well as photoionisation rate) 
with increasing box size and decreasing resolution.
Full details about how the mean free path is measured from the simulations are given in Appendix~\ref{app:measure_mfp}.

\subsection{Spatial UV background fluctuations in the full radiative transfer simulations with ATON}
\label{IGM_prop}

\begin{figure*}
   \begin{center}
      \includegraphics[width=\textwidth,height=\textheight,keepaspectratio]{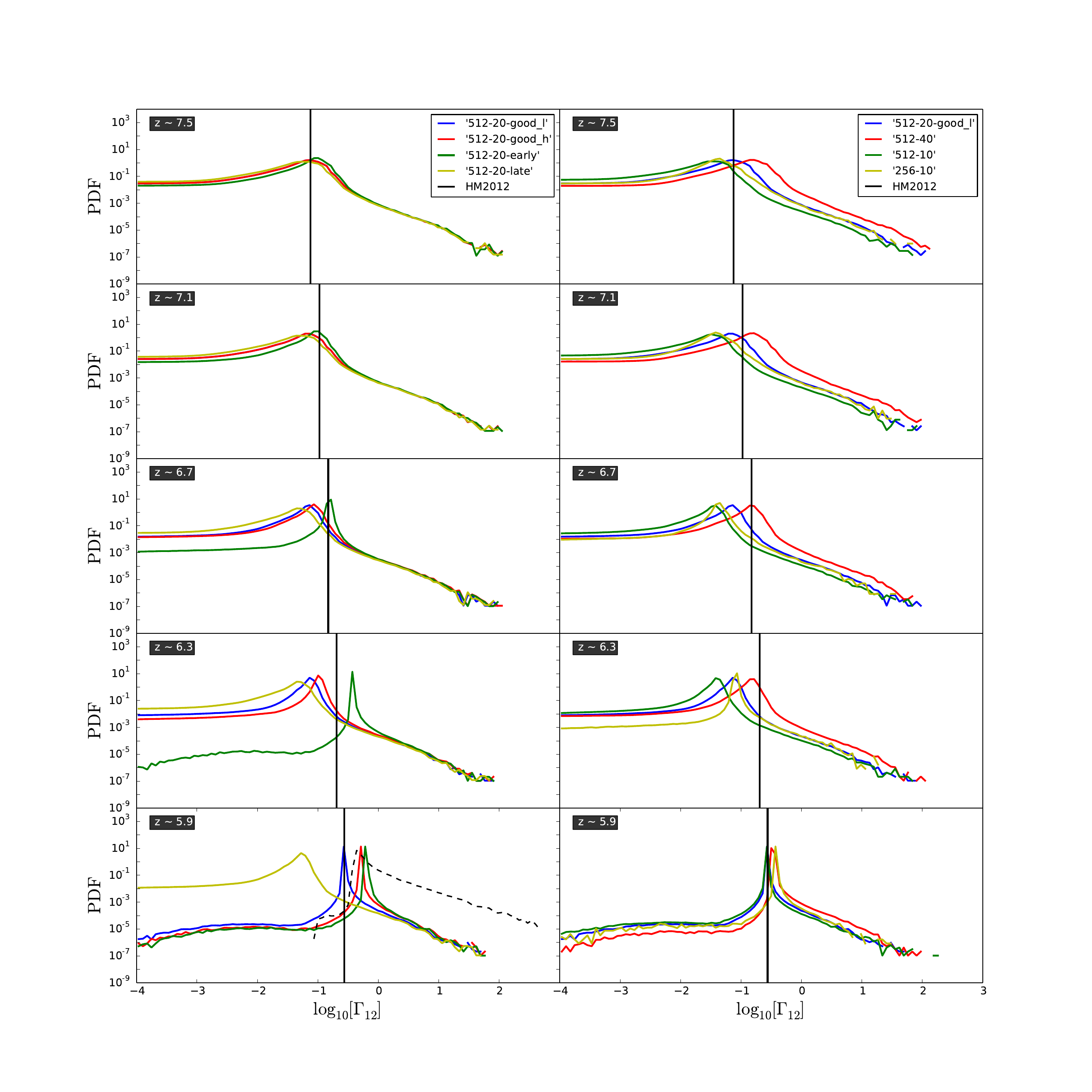}    
  \caption{Probability distribution function (PDF) of the photoionisation rate ($\mathrm{d}P/\mathrm{dlog}_{10}\Gamma_{12}$) at five different redshifts in our  ATON 
simulations with different reionisation histories (left column) and for the simulations with different box sizes and resolutions (right column).  
The solid black vertical lines show the  values of $\Gamma_{12}$ in our 
RAMSES simulations without radiative transfer based on the HM2012 UV background model (their Table 3)  at the same redshifts. 
The dashed black curve in the bottom left panel shows the PDF of our `bright source' model discussed in section \ref{PDF_teff} and Appendix \ref{toymodel}.}
    \label{PDF_gamma_20}
  \end{center}
 \end{figure*}

Here we will have a closer look at the spatial fluctuations of the photoionisation rate in our 
full radiative transfer simulations. 
In Fig.  \ref{PDF_gamma_20}, we show the PDF of the photionisation rate $\Gamma_{12}$ in 
our different ATON simulations for a range of redshifts  as well as the value of $\Gamma_{12}$ of the HM2012 
model used in the RAMSES simulations without radiative transfer  at those redshifts. 

The first thing to note is that, at low redshift, the PDF of the photoionisation rate is strongly peaked and that the spatial fluctuations are rather small. 
With increasing redshift the distributions broadens  considerably  
and a significant tail towards  low values of $\Gamma$ develops. 

As expected the damping of the spatial fluctuations correlates with the timing 
of reionisation and (moderate)  spatial fluctuations persist for some time after full percolation has occured.

The trend of increasing  mean photoionisation rate
with increasing box size and decreasing resolution
is again clearly recognisable. Other possible trends in the PDF of $\Gamma$ with the simulations of different box size and resolution appear  to be weak.

\subsection{Reproducing Lyman-$\alpha$ forest data}

\subsubsection{Extracting mock Lyman-$\alpha$ forest spectra}
\label{spectra_extraction}

We have produced  mock  absorption spectra from our simulations for 
comparison with \lya forest data as e.g. described in \citealt{1998MNRAS.301..478T}. 
For this we  used the gas density, gas temperature, gas velocity and ionisation state of
the gas in the simulation boxes and compute Lyman-$\alpha$ absorption spectra along random lines of sight.
We did this for both the RAMSES simulations without  the radiative transfer and the ATON simulations.
Note again that the temperatures of the (optically thin) RAMSES simulation were used throughout as 
discussed in section \ref{Simulation_set}.
We have extracted 5000 spectra from each  simulation output.
For comparison with the observed  \lya opacity PDF we concatenated simulated spectra to a length covering  
a comoving  distance  of 50 Mpc/h, the length chosen by \citet{2015MNRAS.447.3402B} for their measurements.

\subsubsection{Effective optical depth}
\label{effective_tau}

\begin{figure*}
   \begin{center}
    \begin{tabular}{cc}
      \includegraphics[width=8cm,height=7cm]{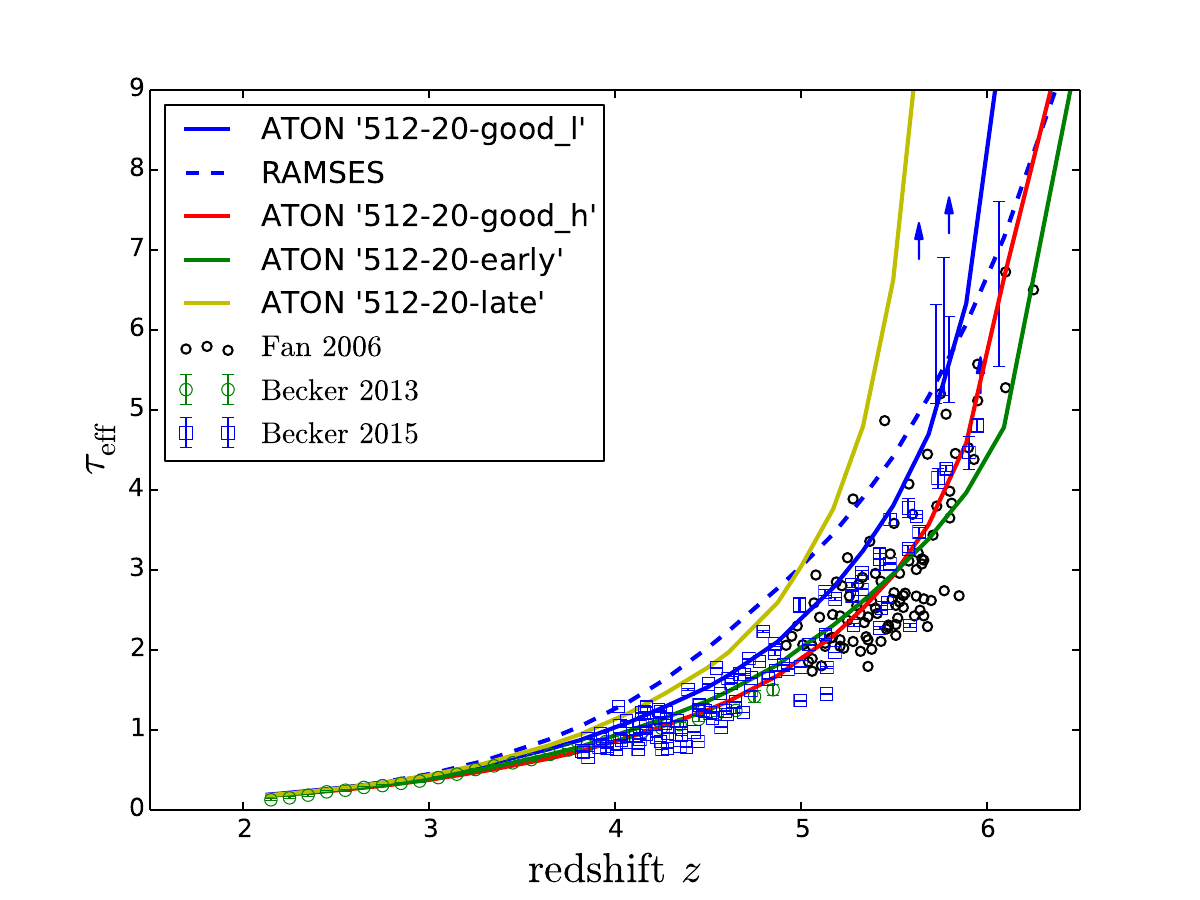} &
      \includegraphics[width=8cm,height=7cm]{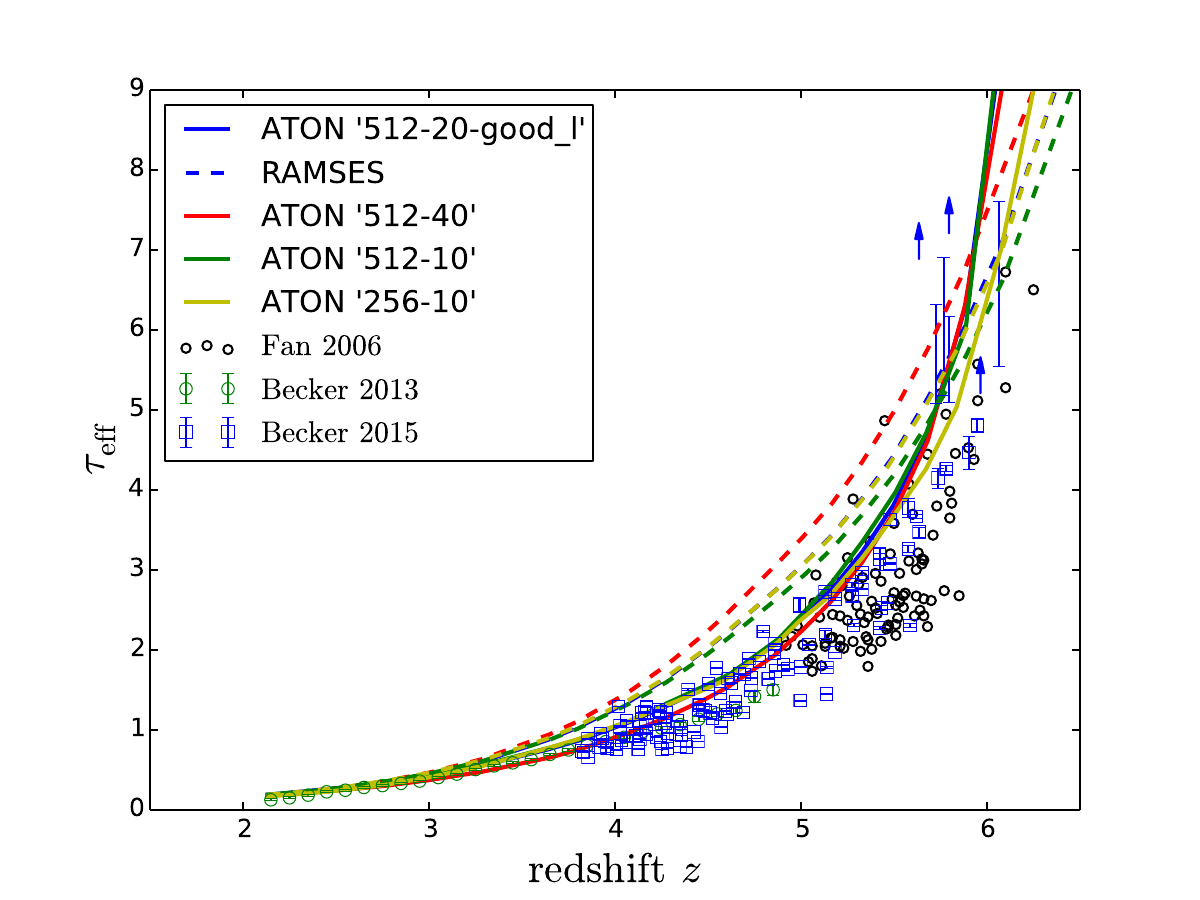} \\
      (a)  & (b)\\
\end{tabular}    
  \caption{(
  (a) The coloured solid curves show the evolution of the of effective Ly$\alpha$  optical depth $\mathrm{\tau_{eff}(z)}$ for the ATON simulations with the different 
       ionisation histories, (b) same as (a) for the simulations with different box sizes and resolutions.
       The  dashed curves  corresponds to the evolution of $\mathrm{\tau_{eff}(z)}$ for the RAMSES simulations 
        without radiative transfer.
	The black data points show the observed $\mathrm{\tau_{eff}}$  for the  QSOs  presented in \citealt{2006AJ....132..117F},
        the green data points show  observational constraints from  \citet{2013MNRAS.430.2067B} and the blue data 
         points show the observed $\mathrm{\tau_{eff}}$  for the  QSOs  presented in \citet{2015MNRAS.447.3402B}. 
}
    \label{teff_vs_z}
  \end{center}
 \end{figure*}

We expect our simulations to reproduce the properties of the $2<z<4$ \lya forest data reasonable well 
as our optically thin $512^3$ RAMSES simulations are very similar to P-GADGET3 simulations that 
some of us have compared extensively to \lya forest data in the past.

In Fig. \ref{teff_vs_z}, we compare the evolution of the effective optical depth $\mathrm{\tau_{eff}}(z)$ 
calculated from 5000 spectra for each of our simulations with  a range of  measurements of the effective optical depth as described in the figure caption.  
Both `good' models reproduce the rapid rise of the observed optical depth 
at $z\sim 5.5$  very well, while in the 'early' and 'late' models reionisation is  completed indeed somewhat 
early and late, respectively.  The effective optical depth of  our `good\_h' model goes thereby right through 
the middle of the observed measurements of $\tau_{\rm eff}$ which however show a large scatter.  The effective optical 
depth of the  `good\_l' model   on the other hand goes through the upper envelope of observed measurements at $5<z<6$
as expected from  our comparison of simulated  photoionisation rates and those 
inferred from \lya forest data in section \ref{gamma_mfp_vs_z}.  Note that our full radiative transfer simulations 
with the rescaled emissivities compared to the HM2012 reproduce the observed  $\tau_{\rm eff}$ 
significantly better than the optically thin RAMSES simulations with the HM2012 model. 
The latter somewhat overproduces the observed $\tau_{\rm eff}$ at $4<z<6$ and 
shows a somewhat slower evolution with redshift than observed. This confirms similar findings by  
\citet{2015arXiv1410.1531P} based on optically thin P-GADGET3 simulations. 
The ionising emissivities  in the HM2012  UV background model  at $5<z<6$ appear to be somewhat 
lower than necessary to reproduce the \lya forest data.  
Judging from Fig. \ref{emissivity_calibration_1} the rather rapid increase of the ionising emissivity at $3<z<6$ due to QSOs appears at fault here.   
An increased contribution from faint AGN at $z>4$ (e.g. \citealt{2015arXiv150202562G}) could  provide a redshift evolution in better agreement 
with the opacity data.

\subsubsection{The incidence of Lyman limit systems}
\label{absorbers}

\begin{figure*}
   \begin{center}
    \begin{tabular}{cc}
      \includegraphics[width=8cm,height=7cm]{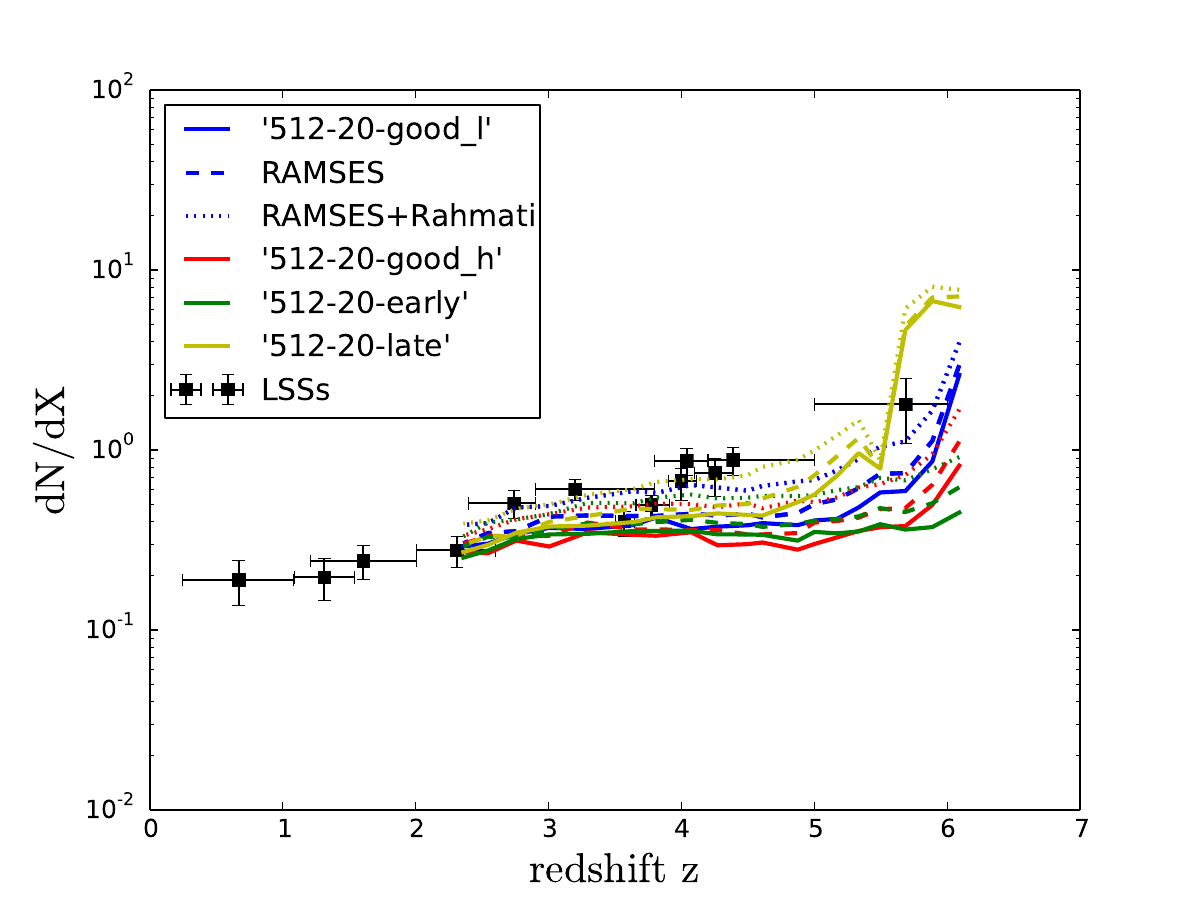} &
     \includegraphics[width=8cm,height=7cm]{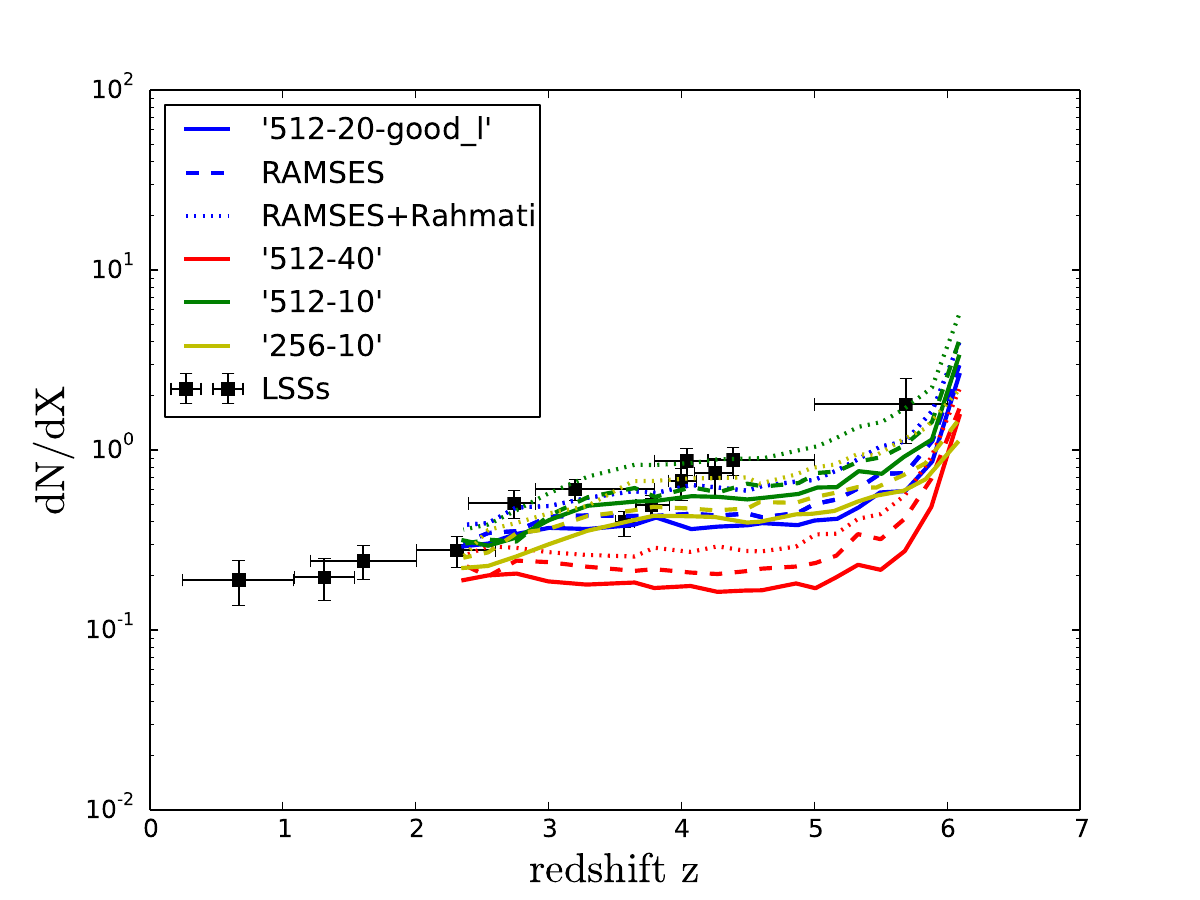}  \\
      (a)  & (b)\\
\end{tabular}    
  \caption{(a) The coloured solid curves show the evolution of the incidence rate 
  dN/dX of LLSs 
for the ATON simulations with the different ionisation histories. (b) same as (a) for the simulations with different box sizes and resolutions.
The  dashed curves  are for the  RAMSES simulations and the dotted curves are for the RAMSES simulations post-processed with the self-shielding correction
of \citet{2013MNRAS.430.2427R}. 
For the RAMSES and RAMSES with self-shielding correction simulations, the neutral fraction of the gas has been rescaled to be consistent with 
the mean $\Gamma$ value of the corresponding ATON simulations.
The black data points with errorbars show the measured  incidence rate of LLSs (from \citealt{2010ApJ...721.1448S} 
and \citealt{2013ApJ...775...78F}).}
    \label{incidence_rate_abs}
  \end{center}
 \end{figure*}

Another important test of the (residual) neutral hydrogen distribution regulating  the \lya opacity 
are  Lyman Limit  Systems  in QSO absorption spectra. For this purpose, we calculated the evolution
of the incidence rate of  absorbers $dN/dX$  where $X$ is the usual absorption distance, 

\begin{equation}
 X=\int_0^z \frac{H_0}{H(z')}(1+z')^2dz'.
\end{equation}

\noindent
and $H(z)$ is the Hubble constant at redshift $z$ (\citealt{1969ApJ...156L...7B}).

In practice we search in our mock spectra for all  optically thick pixels  ($\tau > 3$). 
We compute then the column density of each connected absorber defined in that way
and select those  with $N_{HI} \ge10^{17} \, \mathrm{cm^{-2}}$ as Lyman Limit Systems (LLSs).
We thereby exclude regions not yet ionised where the occurrence of LLSs is
poorly defined. 

In Fig. \ref{incidence_rate_abs} we compare the evolution of $dN/dX$ for all  our simulations
to the observed incidence rate from a cumulation of \lya forest data as described in the figure caption.
The solid curves show the results for the ATON simulations while the dashed and dotted curves are respectively for the optically thin 
simulations without and with the \citet{2013MNRAS.430.2427R} self-shielding correction 
(see also  \citealt{2007ApJ...655..685K} and \citealt{2011ApJ...743...82M}). 

Overall the simulations reproduce the increase of the incidence rate of LLSs with increasing redshift very well, 
but  underpredict the observed  incidence rate  by about a factor 1.5-2 for the  ATON simulations as well as for the RAMSES simulations without radiative transfer.  
This is similar to what has been found by other studies in the 
literature. In optically thin simulations without (supernovae) feedback the neutral hydrogen 
distribution in galactic haloes is generally found to be  too spatially concentrated 
to reproduce the observed LLS incidence rate (\citealt{2010MNRAS.402.1536S,2013MNRAS.436.2689A,2014MNRAS.445.2313B}).

\citet{2011MNRAS.418.1796F} used  high-resolution zoom simulations to
demonstrate that local stellar radiation further reduces    the abundance of high column density absorbers.  
Similar conclusions have been reached by  \citet{2013MNRAS.431.2261R} 
who found that radiative transfer  simulations with local stellar sources  underpredict the rates of incidence
of absorbers with column densities  $10^{19}< N_{HI}<10^{21}\mathrm{cm^{-2}}$  by factors of a few.

Simulations with the same resolution but different box size (i.e. the 512-20-good and the 256-10 models) show very similar evolution.  
There is, however, a  clear increase of the incidence rate of LLSs with increasing resolution at $z>2.5$ and our highest resolution 
512-10 model starts to match the data reasonably  well.  Part of the discrepancy between observed and simulated LLS incidence rate is 
therefore clearly due to insufficient resolution and not the lack of stellar feedback in our simulations.   
Note that our RAMSES simulation with the Rahmati self-shielding prescription tend to give slightly larger 
LLS incidence rates than our full radiative transfer simulations with ATON where ionising sources 
are located within the LLS.

\subsubsection{The cumulative \lya opacity PDF}
\label{PDF_teff}

\begin{figure*}
   \begin{center}
      \includegraphics[width=\textwidth,height=\textheight,keepaspectratio]{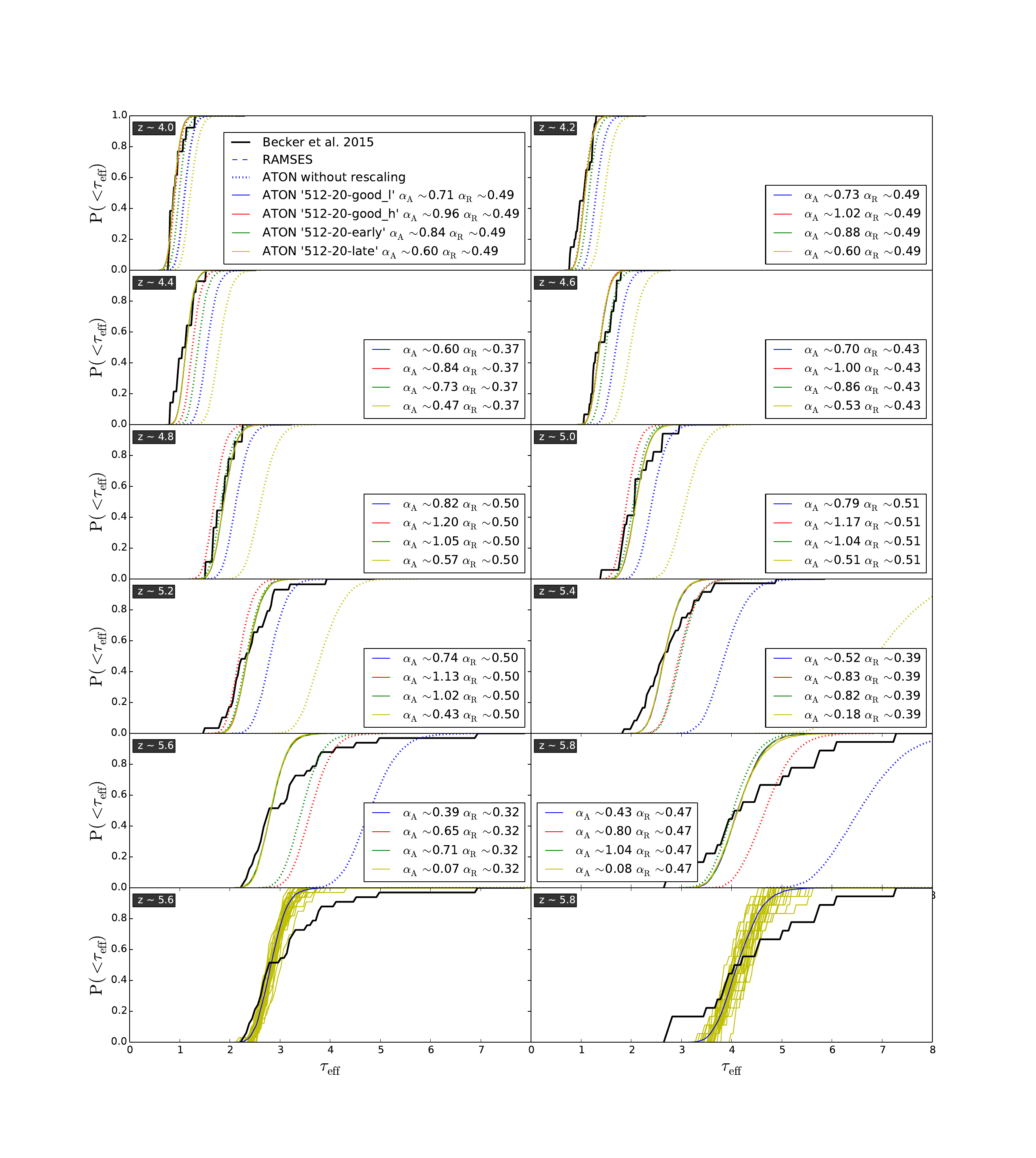}
  \caption{The coloured solid curves show the cumulative PDF of the effective optical depth $\mathrm{\tau_{eff}}$ for our  
  ATON simulations  at ten different  redshifts as indicated on the plot.
The effective optical depths are computed for chunks of 50 comoving Mpc/h as in the observed sample.
The dashed curves show  the corresponding distributions for the  RAMSES simulations without radiative transfer. 
The $\alpha_{A}$ and $\alpha_{R}$ values in the labels of the different panels  are the factors 
by which we have rescaled the optical depth of the ATON and RAMSES  simulations to  match the measurements of \citet{2015MNRAS.447.3402B} shown by the solid black 
curves  at $P(\tau_{\rm eff})=0.5$.
The dotted curves show the corresponding distributions of  the ATON simulations before the rescaling.
The two bottom panels show 30 individual realisations (solid yellow curves) of the cumulative flux PDF for the two 
highest redshift bins for samples of simulated spectra from the 512-20-good\_l ATON simulation with total path length equal to that 
of the observed sample.}
    \label{cumulative_teff_10}
  \end{center}
 \end{figure*}

\begin{figure*}
   \begin{center}
      \includegraphics[width=\textwidth,height=\textheight,keepaspectratio]{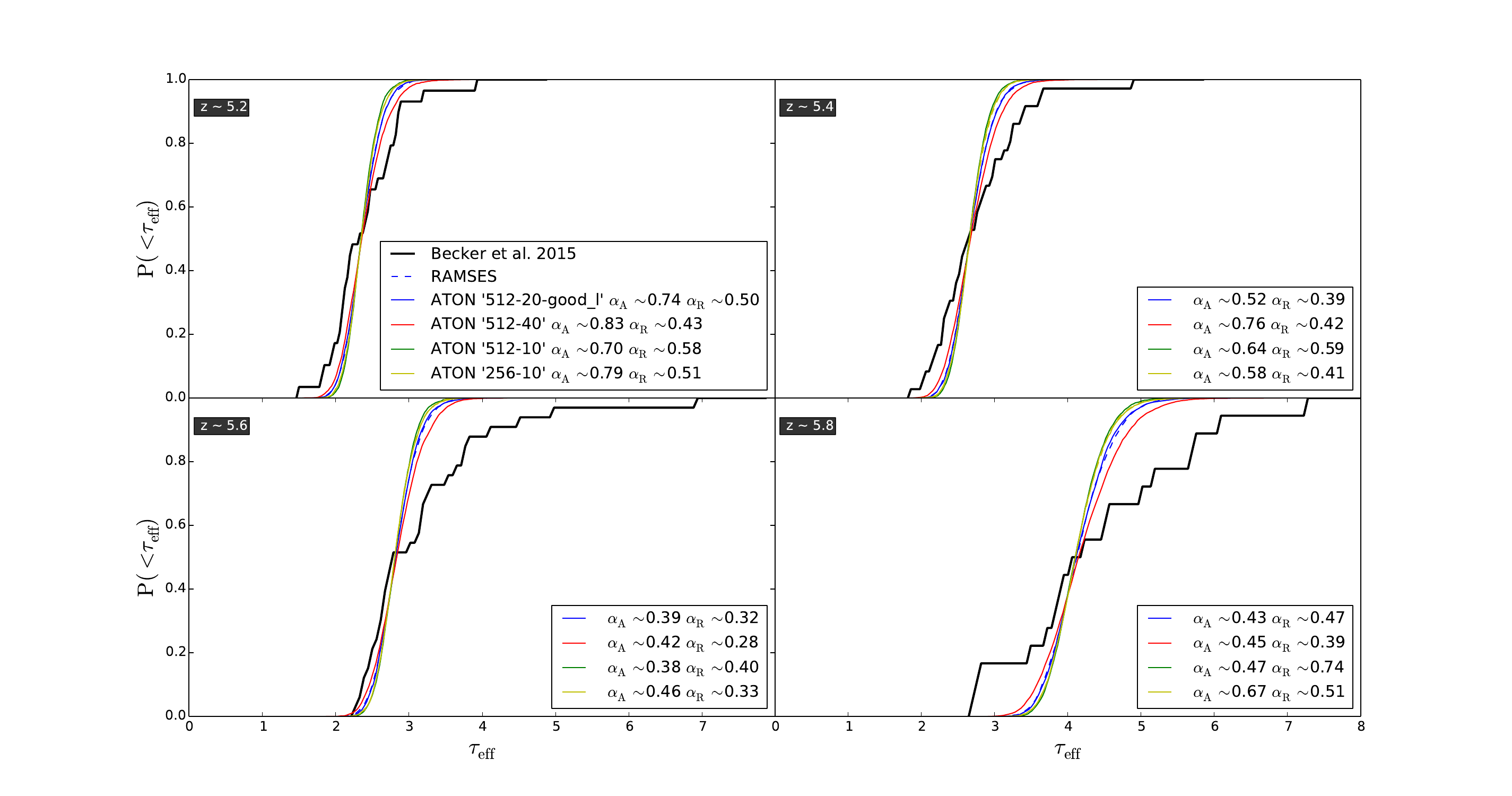}    
  \caption{Same as figure \ref{cumulative_teff_10} for the four highest redshift bins for simulation of different 
  box size and resolution.}
    \label{cumulative_teff_20}
  \end{center}
 \end{figure*}

\begin{figure*}
   \begin{center}
      \includegraphics[width=\textwidth,height=\textheight,keepaspectratio]{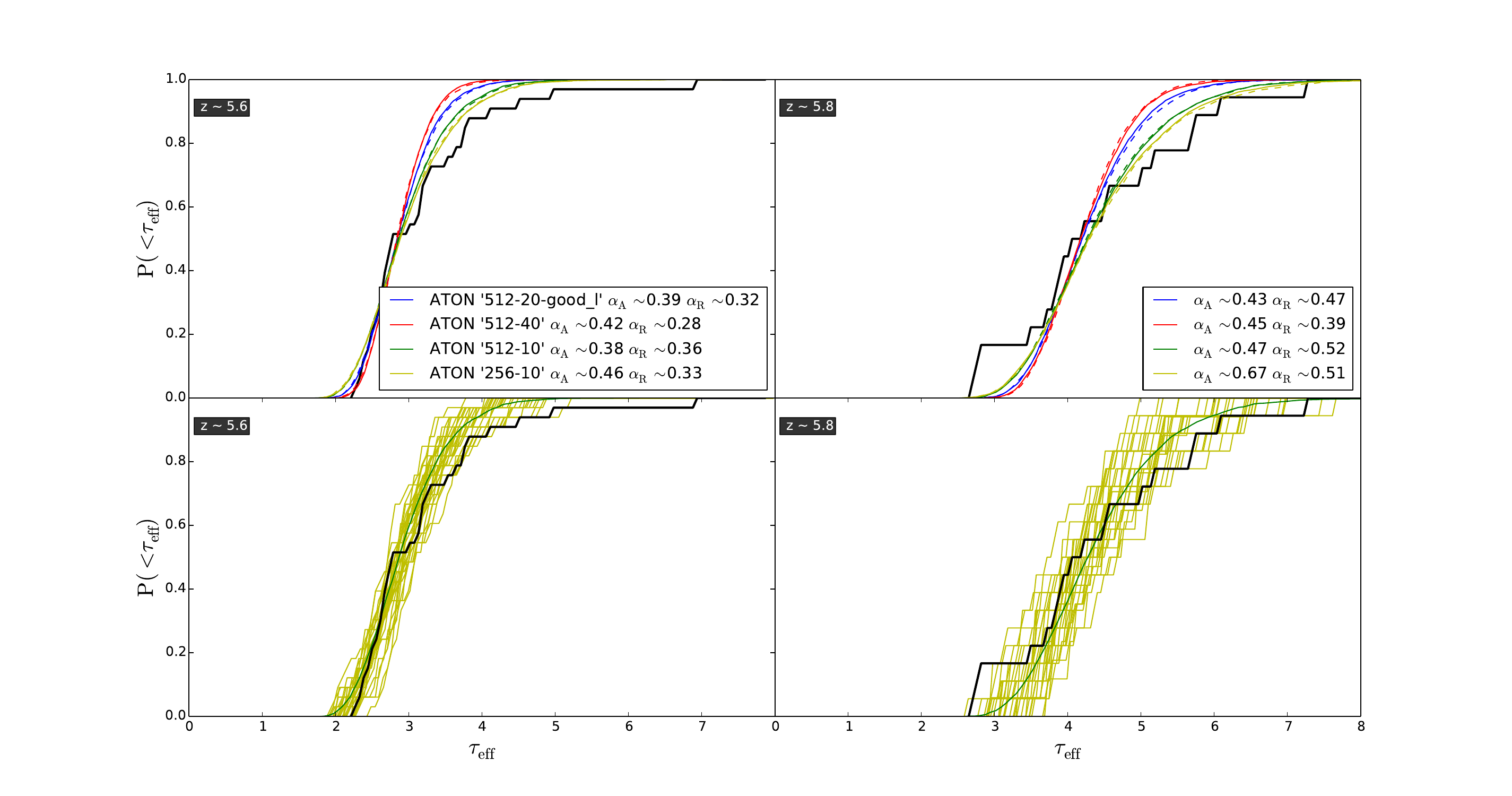}
  \caption{Upper panel : the coloured solid curves show the cumulative PDF of the effective optical depth $\mathrm{\tau_{eff}}$ for our  
  ATON simulations for the two highest redshifts as indicated on the plot.
The effective optical depth is now computed for  chunks of a length  corresponding to the box size in the different simulations instead of chunks of length 50 Mpc/h as before.
The two bottom panels show 30 individual realisations (solid yellow curves) of the cumulative flux PDF for the two 
highest redshift bins for samples of simulated spectra from the '512-10' ATON simulation with total path length equal to that 
of the observed sample. The effective optical depth is again computed for chunks equal to the box size of the simulation, 10 Mpc/h in that particular case.}
    \label{cumulative_teff_box_size_chunk}
  \end{center}
 \end{figure*}

\begin{figure}
   \begin{center}
      \includegraphics[width=9cm,height=7cm]{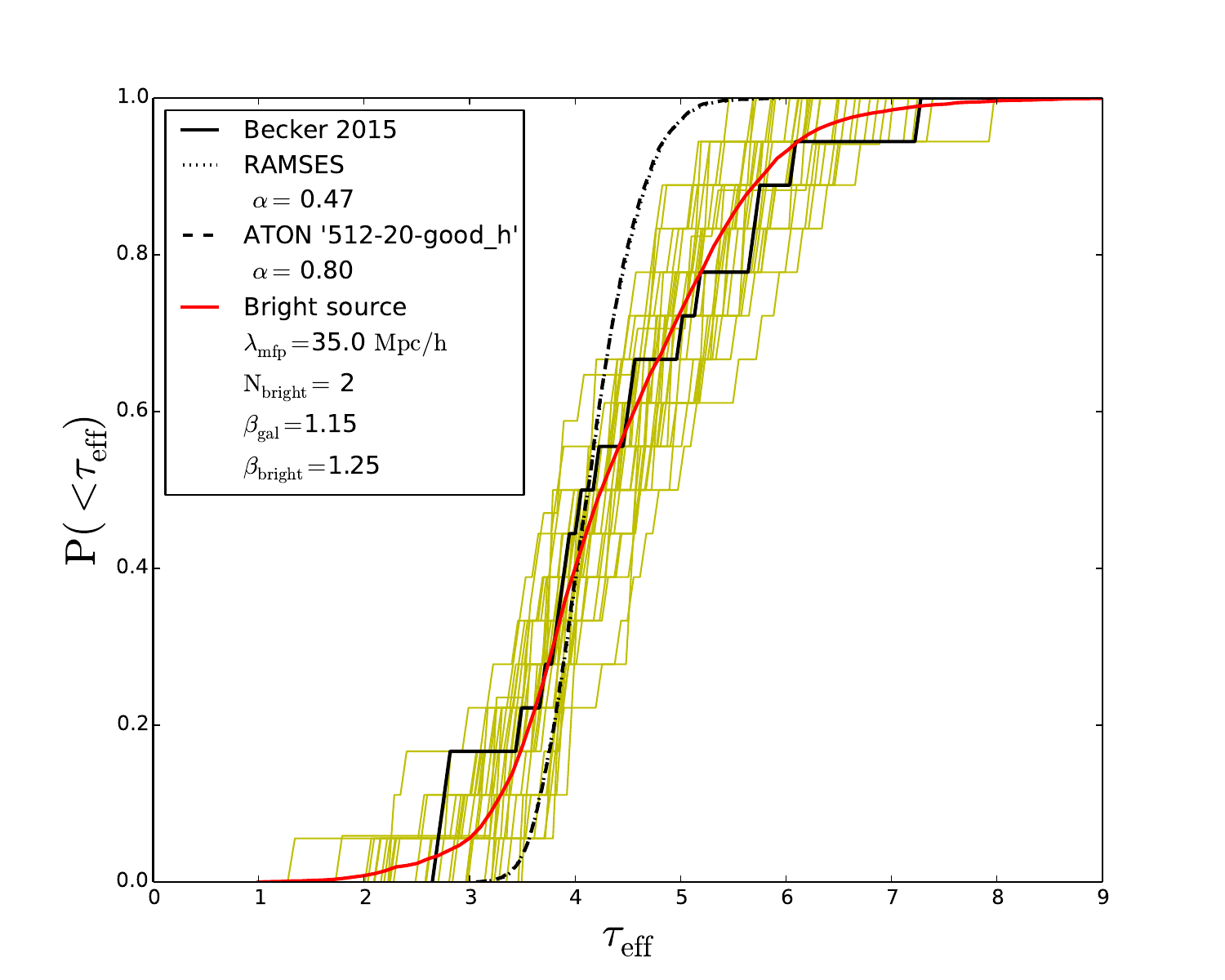}
  \caption{The red solid curve shows  the cumulative  effective optical depth PDF for a ``bright source model" with  two bright ionising sources in a  periodically replicated volume 
of $(100 {\rm Mpc/h})^{3}$ contributing  significantly to the ionising UV background at redshift $z \sim 5.8$.   
The solid yellow curves show 30 individual realisations of the cumulative flux PDF for samples of simulated spectra with photoionisation rates of  
the toy model with total path length equal to that of the observed sample.  The assumed contribution from rare bright sources is added to that in 
our  512-20-good\_l ATON simulation such that the mean photoionisation rate is equal to that in the corresponding good\_h model.
The photoionisation rates are then scaled by a global factor 0.87 to  match the observed PDF
as measured by  \citet{2015MNRAS.447.3402B}. An average mean free path of $\sim 35$ comoving Mpc/h as in our  512-20-good\_h ATON model is assumed   for the attenuation of the bright sources. 
The black dotted and dashed curves show the PDF of the  512-20-good\_h ATON simulation  and the 
PDF of  our corresponding optically thin RAMSES simulation with optical depths  rescaled 
to match the observed PDF at $P(<\tau_{\rm eff}) =0.5$. For more details see text and appendix \ref{toymodel}.
The black solid curve shows  the data from \citet{2015MNRAS.447.3402B}.
}
    \label{cumulative_teff_toy_model}
  \end{center}
 \end{figure}

Recently, \citet{2015MNRAS.447.3402B} have presented accurate measurements of the 
PDF of the \lya effective optical depth obtained from chunks of spectra of 50 comoving Mpc/h at $4<z<6$ based on a large sample of high-quality 
high-redshift QSO absorption spectra. Based on a comparison with P-GADGET3  simulations without radiative transfer similar to our RAMSES simulations 
\citet{2015MNRAS.447.3402B} argued that at $z>5$  the \lya opacity PDF  deviates increasingly from that expected for a spatially homogeneous 
UV background and attributed the increasing deviation at the high opacity end to the spatial UV fluctuations expected  to persist for 
some time during the post-overlap phase of hydrogen reionisation. As we have discussed in section 3.2
our full radiative transfer simulations with ATON show such spatial fluctuations, albeit  at a rather moderate level.

In Fig.  \ref{cumulative_teff_10}  and  \ref{cumulative_teff_20} we compare  the 
cumulative \lya opacity PDF  at $4<z<6$  for our ATON and  RAMSES simulation without radiative transfer with  the measurements of   
\citet{2015MNRAS.447.3402B}.  
As discussed  in section \ref{effective_tau} our ATON simulations for the `good\_h' model, shown here as the red dotted curves, 
reproduce  the mean  effective optical depth  best. 
At high redshift the observed effective optical depth at a given redshift shows, however, clearly a much larger scatter than 
in our simulations. The discrepancy is similar to that found   by \citet{2015MNRAS.447.3402B},
for their optically thin simulations. 

To investigate this further we adjust the  optical depth of our simulations by a constant factor such that 
the mean effective optical depth of  the simulated  spectra is  equal to  $\tau_{\rm eff}$ of  the observed distribution at $P(\tau_{\rm eff})=0.5$.
The correction factor $\alpha$ by which the optical depth is multiplied is indicated  for each simulation on the plots. 
Let us now  again have a  look at  Fig. \ref{cumulative_teff_10}  where we compare the different 
reionisation histories. At $z<4.8$ the \lya opacity PDF from the ATON and optically thin RAMSES simulations 
are almost identical after rescaling. They also  agree very well  with the observed PDF,  very similar to what was found by  \citet{2015MNRAS.447.3402B} based on their P-GADGET3 simulations.  
At $z \sim 5$ and $z\sim 5.2$  the PDF of the simulated spectra 
begins to exhibit  a somewhat steeper slope than the observed PDF.  At $z>5.4$ the  discrepancy 
increases and especially in the two highest redshift bins observed and simulated PDF of the 
effective optical depth are strikingly different. 

The observed sample is still rather small. In order to get a feel for the expected 
sample variations  the two bottom panels of Fig. \ref{cumulative_teff_10} show, for the two highest redshift bins, the  observed \lya 
opacity PDF with 30 individual  realisations for samples of simulated spectra of the `520-20-good\_h' model   
with total path length equal to that of the observed sample.  The scatter is substantial, but the discrepancy between observed  
and simulated PDF is clearly much larger than can be explained with sample to sample 
variations.

In Fig. \ref{cumulative_teff_20},  we show the \lya opacity PDF for simulations where we have varied box size and resolution.   
The differences  are  moderate. As expected increasing the box size somewhat increases the 
width of the distribution, but not nearly enough to attribute  the discrepancy with the observed PDF to 
an insufficient  box size of our simulations. 

That  the PDF of the effective optical depth of our  ATON and RAMSES simulations are rather similar 
means that the spatial UV fluctuations in our full radiative simulations are not sufficient to explain the 
observed rather wide distribution of optical depth at $5<z<6$.  \citet{2015MNRAS.447.3402B} 
modelled  the expected UV fluctuations in the post-overlap phase for 
ionising emission from (faint) galaxies.  
Combining  a simple model for the attenuation of ionising photons for  plausible assumptions for the mean 
free path  with their optically thin simulations they  came to similar conclusions.  

\citet{2015MNRAS.447.3402B}  therefore proposed that the increasing tail 
of large optical depth in their data could be due to large spatial  fluctuations of the mean free
path of ionising photons expected in the post-overlap phase of reionisation  (Furlanetto \& Oh (2005), Mesinger \& Furlanetto 2009).

To shed further light on this  in Fig. \ref{cumulative_teff_box_size_chunk}, we look at the influence of the chunk size used to compute the effective optical depth. We use our simulations of different
box sizes for this and set the chunk size equal to the size of the simulation box.  
Reducing the length of the spectrum used to compute the effective optical depth  increases the scatter in $\tau_{\rm eff}$ 
and thus the width of the PDF as expected.  Interestingly for a chunk/simulation box  size of 10 Mpc/h 
the optical depth PDF matches the observed PDF (calculated with a chunk size corresponding to 50 Mpc/h)
to within the expected sample variation. Our simulations  appear to reproduce the observed range 
of observed optical depth at a given redshift, but the coherence length of the optical depth appears 
to be much shorter than observed so that averaging reduces the width  of the PDF.

Density fluctuations on these rather large scales are very small.
However, as we will show below large scale spatial UV fluctuations that extend over 50 Mpc/h 
scales or more could explain this large coherence length.   As we and  \citet{2015MNRAS.447.3402B}  
have shown, such fluctuations   are, however, at odds with a UV background at $5<z<6$ dominated by the large 
number of faint galaxies widely believed to be responsible for reionisation. Leaving this point aside here
for the moment (we will come back to this in the next section) we explore here a (toy) model (see also Appendix \ref{toymodel}), 
where the galaxies driving reionisation contribute only a fraction of the UV background at $5<z<6$, while the remainder  
is contributed by much rarer sources (faint AGN or perhaps very bright galaxies).

Figure \ref{cumulative_teff_toy_model} shows the  resulting effective optical depth PDF for  a model with  two bright sources in a 
(periodically reproduced) cube with  side-length  100 Mpc/h. We have used  the photoionisation rates in our   
`512-20-good\_l' model    and have added the UV emission from the rare bright sources such as to 
reproduce the mean $\Gamma$ of the 512-20-good\_h ATON model which is the closest to the observations. 
The resulting large scale fluctuations of the ionising flux significantly widen the optical depth PDF. 
As Figure \ref{cumulative_teff_toy_model} shows this particular set of parameters reproduces the observed PDF at $z=5.8$ very well.

The large scale fluctuations of the ionising flux  due to the rare sources in this  model were calculated with the simple attenuation 
model first introduced by \cite{2006MNRAS.366.1378B}
and used by \citet{2015MNRAS.447.3402B}   (but for a much higher space density than considered here). 
For details and limitations of this toy model we refer the reader once more to  
Appendix \ref{toymodel}. When choosing the parameters of the  model we found that the optical depth PDF depends very sensitively 
on the space density of sources, mean free path of ionising sources and fraction of the UV background contributed by 
rare  sources and there will certainly be other parameter choices that will also reproduce the data. 
We should thus emphasise here that our toy model should just be seen as proof of concept. 
We will leave a more detailed exploration of the effect of spatial UV fluctuations due to rare bright sources on the effective optical  depth PDF  to a future paper.

\subsection{Implications for the role of bright galaxies and (faint) AGN  for the origin and evolution of the ionising emissivity at high redshift}

As discussed in detail in the previous sections with our ATON simulations we have  got good  agreement  with the current observational constraints 
from \lya forest data in the redshift range  $2<z<6$.
With a moderate adjustment of the evolution of the ionising emissivity compared to HM2012 we  
have been  able to reproduce the observed evolution of the mean neutral fraction, mean photoionisation rate, averaged mean free path, 
mean \lya opacity and the incidence rate of LLSs.

Similar to results from optically thin simulations we  failed, however, 
with our radiative transfer simulation to reproduce the broad PDF of the optical depth at $5<z< 6$ when measured for chunks of 50 Mpc/h as in 
 \citet{2015MNRAS.447.3402B} . We have thereby  confirmed the result of \citet{2015MNRAS.447.3402B}   that the  corresponding optical depth fluctuations  
on these large scales are much larger than can be plausibly explained as being due to density fluctuations.  Judging from the large  differences in the  
visual appearance of individual QSO spectra, the implied large coherence length of order unity optical depth fluctuations appears  to be even larger than 50Mpc/h.  
The very extended  Gunn-Peterson trough of high \lya optical depth in ULAS J0148+0600 at $z\ge 5.5$ is particularly striking in this regard.

Addressing this properly will  require much larger simulated regions at similar resolution(s) than we have employed here and will be a 
formidable computational task. As  we have demonstrated  with our toy model in the last section a significant contribution to the  
integrated ionising UV  background from  bright sources with space densities of order $\mathrm{10^{-6} ({\rm Mpc/h})^{-3}}$
appears to be required. 

While most of the literature including HM2012 have assumed  that the contribution of
QSOs drops rapidly  at $z>4$, some authors  have argued  that there may be a significant contribution by AGN to the ionising 
UV background at $z>4$ (see \citealt{2015arXiv150202562G} for a recent discussion). As discussed by \cite{2015A&A...575L..16H}  a contribution 
of (faint) AGN at $z>4$ is consistent with the soft X-ray background, but
sources with softer spectra would have to be responsible for driving hydrogen reionisation in order not to exceed 
measurements of the soft X-ray background. In light of the  need for spatial fluctuations of the ionising UV background on scales $\ga 50$ Mpc/h at $5<z<6$ 
it appears therefore prudent to also consider the possibility of efficient escape of ionising photons from stars in rare very bright (possibly star bursting) galaxies at $z>5$. 
These galaxies would, however, have to have unusually hard ionising  spectra  and large (of order unity) escape fractions to provide 
the required ionising emissivity. Interestingly the recent discovery of strong CIV emission in high-redhift galaxies by \citet{2015arXiv150406881S} may point in this direction. 
In either case the current widely accepted assumption that the ionising UV background at $z>5$ is dominated  by faint sub-$L*$ galaxies may  need  revision.

\section{Conclusions  and Outlook}
\label{prospects}

We have calibrated here full cosmological radiative transfer simulations performed with ATON/RAMSES with \lya forest data  and compared them with optically thin RAMSES simulations.

Our main results are as follows.

\begin{itemize}

 \item{Once the amplitude  of ionising emissivities  in the 
 HM2012 UV background model is rescaled to take into account (the resolution dependent)  recombination in the haloes of ionising sources, we were able to obtain reionisation histories 
for our  full radiative transfer simulations in very good 
 agreement with a wide range of available observational constraints derived from \lya forest data: 
 mean effective optical depth, mean free path of ionising photons, incidence of LLSs, photoionisation rates. }

\item{
To reach good agreement with the  rather flat evolution of  inferred photoionisation rates at $2<z<6$ required, however,  a moderate re-scaling  of the evolution of 
the ionising emissivity with redshift compared to the HM2012 model.
The HM2012 model tends to have a too steep redshift evolution of the ionising emissivity in the redshift 
range $3<z<6$, while in the same range its normalisation is too low to produce the  photoionisation rates inferred from \lya forest data.}

\item{In simulations that match the \lya forest data well, percolation of individual HII regions occurs at $z\sim7$ and   moderate spatial fluctuations of the photoionisation rate persist to about $z\sim 6$.
Reionisation histories where percolation  occurs sufficiently late to allow for larger spatial fluctuations of the photoionisation rate  
at $z<6$ appear to be inconsistent with the \lya forest data.}
 
\item{Reproducing the broad PDF of the effective optical depth in chunks of 50 comoving Mpc/h at redshift $z\sim 5-6$ proved rather difficult. 
Spatial UV fluctuations on much larger scales  ($\ga$ 50 comoving Mpc/h) than produced 
by (sub-) $L_*$ high-redshift galaxies  appear to be required to reproduce the observed PDF of the optical depth suggesting a contribution  of order unity to the ionising  
UV background by sources with space densities of order $10^{-6} ({\rm Mpc/h})^{-3}$. This is very different from the  widely accepted assumption made in HM2012 that 
the ionising emissivity at these redshifts is dominated by very faint sources, but is consistent with recent suggestions for a significant contribution of AGN  
to the ionising emissivity at $z>4$.  
Such an increased contribution from faint AGN at $z>4$ could at the same time  provide a redshift evolution of the ionising emissivity  
in better agreement with the redshift evolution of opacity data.
Better characterising the coherence length of the spatial fluctuations of  the \lya optical depth at $z>5$  should  provide valuable further
information on the contribution of rare bright sources  to the ionising emissivity at high redshift.} 

 \end{itemize}
 
Further progress will require full radiative transfer simulations of much larger regions at a similar resolution as that of the 512-20 models   discussed here that include rare bright galaxies and AGN. 
Another important prerequisite for further progress  will be the realistic modelling  of the angular  and temporal distribution of the ionising radiation from starbursts and AGN including light travel time effects. 
Such simulations of larger regions  should hopefully  enable us to  obtain realistic predictions of the (coupled) spatial fluctuations of mean-free path  of ionising photons and photoionisation rates on the relevant large spatial scales. 
It will also be important to further improve the modelling of the gas distribution in galactic haloes by including stellar feedback in order  to better reproduce the incidence of LLSs acting as sinks of ionising radiation.

Other avenues for improvement are multi-frequency treatment of the radiative transfer with the aim of an improved  modelling of the temperature evolution of the IGM during hydrogen and helium reionisation,   
the self-consistent 
coupling of  the effects of radiation pressure and the photoheating of the gas and eventually a  self-consistent modelling of 
the ionising sources including possible self-regulating effects due to the  reionisation of hydrogen. 

Overall we have shown here that \lya forest data provides very strong constraints on how reionisation proceeeds
and we expect   \lya forest data  to become  an even more important calibrator of further improved full radiative transfer simulations  
of the reionisation of hydrogen as we push to higher redshift with ongoing and planned QSO surveys.

\section*{Acknowledgments}

We thank George Becker for  making his measurements of the \lya opacity PDF available before publication 
and James Bolton and George Becker for giving very helpful advice and comments on the manuscript. This work was supported by the ERC Advanced Grant 320596
``The Emergence of Structure during the epoch of Reionisation". 
The RAMSES simulation were performed utilizing the supercomputer COSMOS Shared Memory system at DAMTP, University of Cambridge operated on behalf of the STFC DiRAC HPC Facility. 
This equipment is funded by BIS National E-infrastructure capital grant ST/J005673/1 and STFC grants ST/H008586/1, ST/K00333X/1.
The ATON radiative transfer simulation
in this work were performed using the Wilkes GPU cluster at the University of Cambridge High Performance Computing Service (http://www.hpc.cam.ac.uk/), 
provided by Dell Inc., NVIDIA and Mellanox, and part funded by STFC with industrial sponsorship from Rolls Royce and Mitsubishi Heavy Industries.
DA is supported by the Agence Nationale de la Recherche Grant ANR-12-JS05-0001 ``EMMA''.

\bibliographystyle{mn2e}
\bibliography{biblio}

\appendix

\section{The halo mass funtion in the RAMSES simulations}
\label{halo_mass_func}

\begin{figure*}
   \begin{center}
      \includegraphics[width=\textwidth,height=\textheight,keepaspectratio]{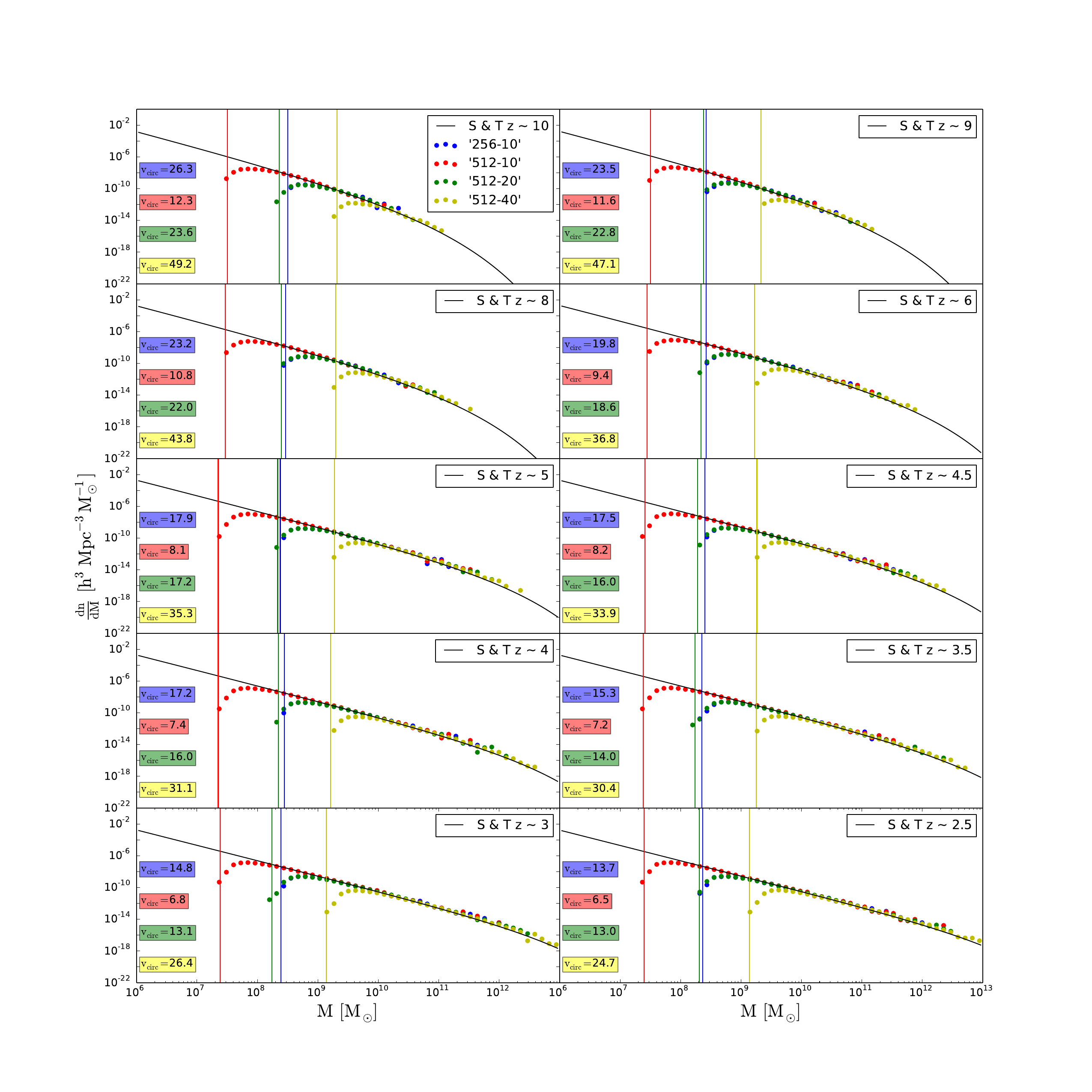}    
  \caption{Evolution of the halo mass function in our different simulations.
The analytical model of \citet{2001MNRAS.323....1S} is shown by the black solid curves.
The coloured vertical lines show the minimum mass assumed to host ionising sources in each model. 
The corresponding halo minimum circular velocities are displayed on the plot (in km/s). 
}
    \label{HMF_10redshifts}
  \end{center}
 \end{figure*}

We have identified the dark matter haloes in our RAMSES simulations with the sources of ionising photons in  our ATON simulations. In this appendix, 
we present the evolution of the halo mass function of our RAMSES simulations with different resolution and box sizes (Fig. \ref{HMF_10redshifts}).
For comparison, we also plot the widely used analytical fit of  \citet{2001MNRAS.323....1S} at each redshift
which has been tested against a wide range of numerical simulations. The halo mass function  of our different 
simulations are in good  agreement with the analytical model over the mass range where agreement should be
expected. The minimum mass of haloes assumed to host ionising sources is indicated by the vertical dashed 
lines as well as the corresponding circular velocities.

\section{The luminosity function of ionising sources}
\label{LFevolution}

\begin{figure*}
   \begin{center}
      \includegraphics[width=\textwidth,height=\textheight,keepaspectratio]{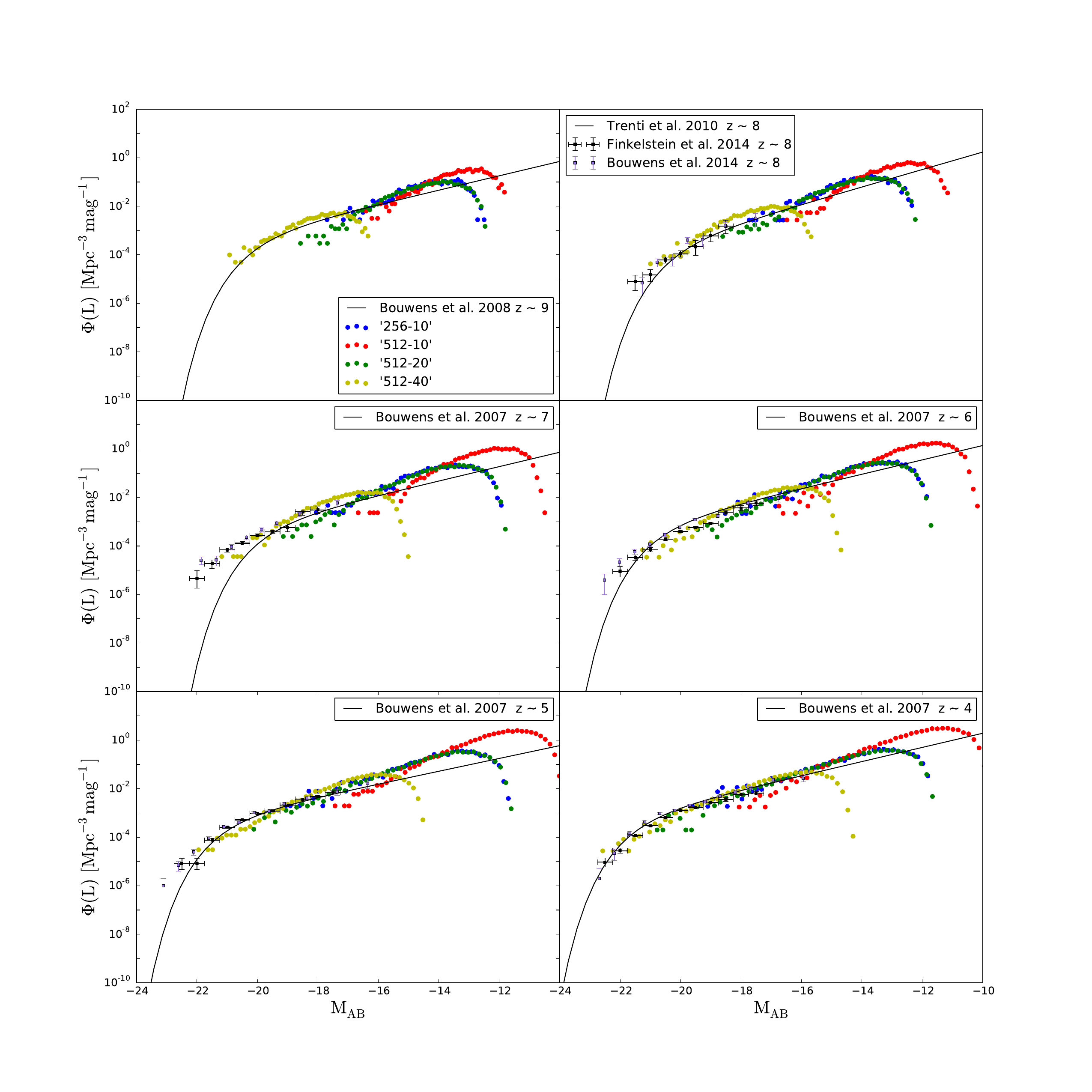}    
  \caption{Evolution of the luminosity function in our different simulations assuming a linear relation 
  between halo mass and 1500 \AA{} flux as described in the text. Black and purple squares with errorbars show respectively 
   observational constraints from \citet{2014arXiv1410.5439F} and \citet{2014arXiv1403.4295B}. The solid curves 
   are fits to observational data from  \citet{2007ApJ...670..928B}, \citet{2008ApJ...686..230B} and \citet{2010ApJ...714L.202T}. 
}
    \label{LF_8redshifts}
  \end{center}
 \end{figure*}

In our ATON  simulation we have assumed that the luminosity of ionising sources scales linearly with the mass of the dark matter haloes identified 
in the RAMSES simulations as discussed in  Appendix \ref{halo_mass_func}.  As discussed by e.g. \citet{2010ApJ...714L.202T} 
models with a linear relation between  dark matter halo mass function and galaxy luminosity fit the evolution of the high-redshift luminosity function reasonably well. 
To test for consistency  of our modelling and relate our assumed population of ionising sources to observed high-redshift galaxies we show here 
the evolution of the luminosity function obtained from such a linear relation between galaxy luminosity and halo mass and 
the halo mass functions discussed in Appendix \ref{halo_mass_func}. 

In our source model, the total comoving ionising emissivity is scaled to that in the 
HM2012 UV background model   as described in section
\ref{photheating}\footnote{The comoving emissivity table of HM2012 (in $\mathrm{erg \, s^{-1} Hz^{-1} Mpc^{-3}}$) can be 
downloaded at {\tt  http://www.ucolick.org/$\sim$pmadau/CUBA.}}.
From this, we can calculate the integrated emissivity  $\epsilon_{1500}^{\mathrm{tot}}(z)$ at 1500 \AA{} at  a given redshift.
Summing  over all ionising sources  gives then the normalization between halo mass 
and 1500 \AA{}  flux $f_{1500}$.

Figure \ref{LF_8redshifts} compares the resulting luminosity functions with  the observed luminosity function of high-redshift galaxies.
At  the highest redshifts with available data  $z\sim 9-8$, there is excellent agreement of our simple model of the luminosity function and the data.
At lower redshifts we  somewhat overpredict the faint end of the luminosity function.
Note, however, that we are  interested in the hydrogen ionising luminosity for our reionisation simulations and that 
the escape fraction of ionising photons is generally believed to  increase with decreasing galaxy luminosity. The somewhat steeper 
luminosity function for the ionising flux we have assumed is therefore probably as good a bet as currently possibly.

\section{Measuring the mean free path of ionising photons from simulations}
\label{app:measure_mfp}

The mean free path of ionising photons in simulations can be computed from the remaining transmission after a specific path length and by assuming 
that the average transmission decreases exponentially with increasing distance. A more sophisticated method is to adjust the length scale over 
which the transmission is measured depending on the value of the mean free path \citep{2013ApJ...763..146E}.

Here we use a more direct approach which does not rely on choosing a length scale for the measurement or on assuming a functional 
form of how the transmission decreases with distance. Instead we directly use the definition of the mean free path,
\begin{equation}
\lambda_{\rm mfp} = \langle \frac{\int x \, \textrm{d}f}{\int \textrm{d}f} \rangle. \label{eq:mfp}
\end{equation}
Here $f(x) = \exp(-\tau(x))$ is the transmitted ionising flux at path length $x$ and $\langle ... \rangle$  denotes an average over many lines of sight. The only assumption we need is that the neutral hydrogen density is constant within each simulation cell.

Without loss of generality we can assume that the initial ionising flux $f(x=0) = 1$. In this case the integral in the denominator in Eq.~(\ref{eq:mfp}) is equal to $-1$, so that we are left with
\begin{equation}
\lambda_{\rm mfp} = -\langle \int_1^0 x \, \textrm{d}f \rangle = \langle \sum_i \lambda_i \rangle,
\end{equation}
where $\lambda_i$ is the contribution of the $i$-th cell in the line of sight to the integral. Next, we note that
\begin{equation}
-\textrm{d} f = e^{-\tau(x)} \, \frac{\textrm{d} \tau}{\textrm{d} x} \, \textrm{d} x = e^{-\tau(x)} \, \frac{\tau_i}{\Delta x} \, \textrm{d} x.
\end{equation}
In the second equality, we have used the assumption that the neutral hydrogen density is constant within each cell so that $\textrm{d} \tau / \textrm{d} x = \tau_i / \Delta x$, were $\tau_i$ is the optical depth of cell $i$ and $\Delta x$ is the cell size. Before performing the integration over cell $i$, we change the integration variable by defining $l \equiv x - (i-1) \Delta x$, which is just the distance from the beginning of cell $i$. Using this we find
\begin{equation}
\tau(x) = \sum_{j=1}^{i-1} \tau_j + \frac{\tau_i}{\Delta x} l,
\end{equation}
so that we can write $\lambda_i$ as
\begin{align}
\lambda_i &= e^{-\sum_{j=1}^{i-1} \tau_j} \int_0^{\Delta x} \left[ (i-1) \Delta x + l \right] e^{-\frac{\tau_i}{\Delta x} l} \frac{\tau_i}{\Delta x} \textrm{d} l \nonumber \\
          &= e^{-\sum_{j=1}^{i-1} \tau_j} \big[ (1-e^{-\tau_i}) (i-1) \Delta x \nonumber \\
          & \quad\quad\quad\quad\quad + \Delta x \frac{1-e^{-\tau_i} (1 + \tau_i)}{\tau_i} \big].
\end{align}
Using this expression, we find the mean free path of ionising photons along a specific line of sight by summinng over the cells along it, i.e. by computing $\sum_i \lambda_i$. 
We perform this summation at least until we have reached the opposite side of the simulation box. 
If the transmitted flux fraction there is larger than $10^{-8}$, we continue the summation along another randomly chosen line of sight. 
We go on in this manner until $f(x) < 10^{-8}$. Finally, we repeat this prodecure many times for different lines of sight and average the obtained mean 
free path values to get a global estimate of the mean free path of ionising photons.

\section{A toy model for rare bright galaxies/(faint) AGN}
\label{toymodel}

\begin{figure*}
   \begin{center}
      \includegraphics[width=\textwidth,height=\textheight,keepaspectratio]{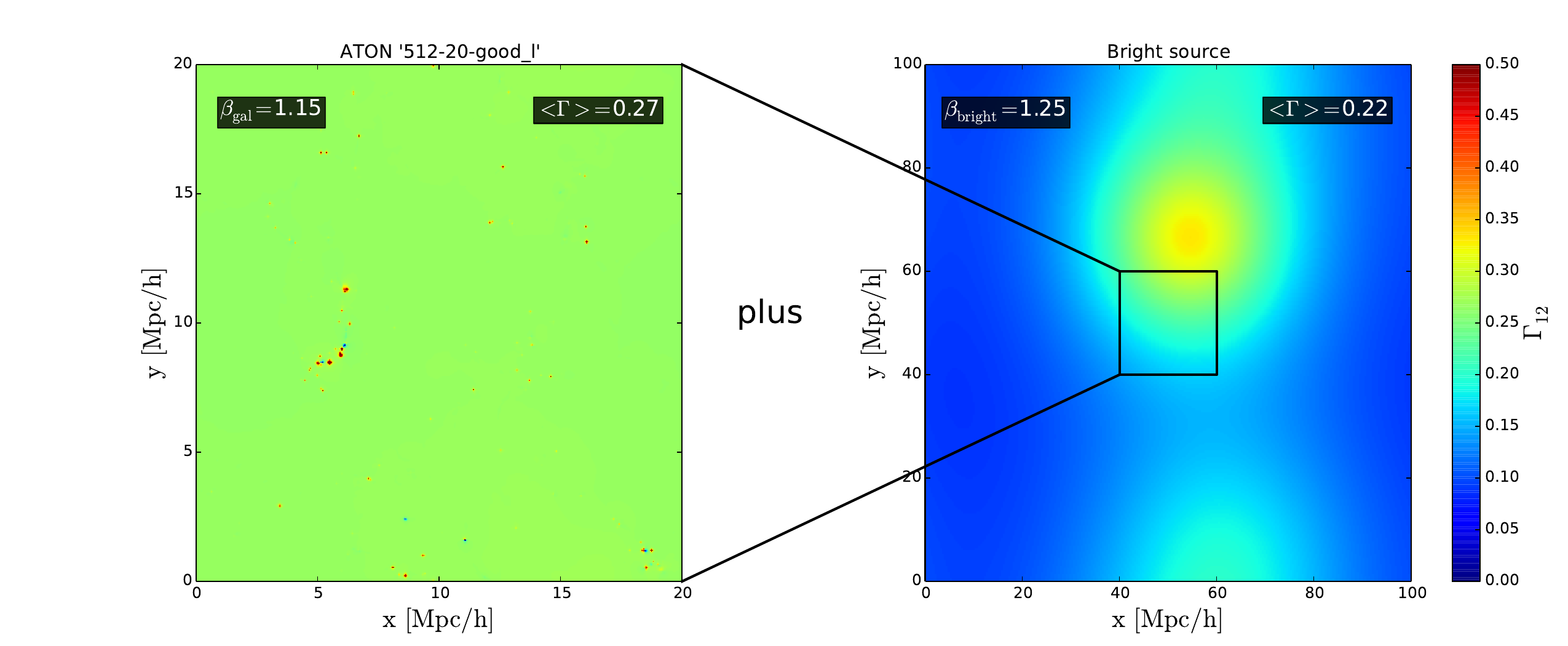}    
  \caption{{\it Right:} The photoionisation rate   $\mathrm{\Gamma_{12}}$   due to  bright sources at redshift $z\sim5.8$ 
  for a slice of our bright source model   assuming a mean free path of $\lambda_{\mathrm{mfp}}^{912} \sim 35$  comoving Mpc/h and 2  bright  
ionising sources in the periodically replicated $(100 {\rm Mpc/h})^3$ volume. The photo ionisation  rates are normalised to match the observed cumulative 
\lya optical depth  as described in the text with $\beta_{\rm gal } =1.15 $ and  $\beta_{\rm bright } =1.25$ and the mean photoionisation rate indicated on the plot is averaged over the whole simulation box. 
 {\it Left:} The photoionisation rate $\Gamma_{12}$  in the  `512-20-good\_l' ATON radiative transfer model, 
also  at redshift $z\sim5.8$. 
}
    \label{gamma_toy_model}
  \end{center}
 \end{figure*}

In section \ref{PDF_teff} we have argued that  fluctuations of the photoionisation rate caused by  rare bright galaxies or perhaps more likely
(faint) AGN may  be responsible of the broad \lya opacity PDF measured for 50 Mpc/h chunks at  $z\sim5.6-5.8$. Here we present details of the  
toy model  discussed  there that we have used to test the effect of rare bright sources  on the \lya opacity PDF at redshift $z\sim5.8$. 

We  use the location  of the most massive dark matter haloes in  a periodically replicated large volume simulation with a box size  100 Mpc/h on a side   
($1200^3$ particles, also  used and described in \citealt{2014arXiv1412.4790C}) at 
$z\sim5.8$ as positions for the bright ionising sources. We then  compute the photoionisation rate $\Gamma$ due to these bright sources at 
every position in the 100 Mpc/h volume with  the simple attenuation model used  by \cite{2015MNRAS.447.3402B} (their galaxy UVB model of section 4.2, 
but see    also \citealt{2011MNRAS.414..241B} and \citealt{2013PhRvD..88d3502V}).

We assume a spectral energy distribution appropriate for AGN for the bright sources of the form  (see \citealt{2001AJ....122..549V} and \citealt{2002ApJ...565..773T}), 
\begin{equation}
L_{\nu} \propto \left\{\begin{array}{ccc}
 \nu^{-0.44} & \mbox{ if $\lambda > 1300 \, \mathrm{\AA}$ }\\
 \nu^{-1.57} & \mbox{ if $\lambda < 1300 \, \mathrm{\AA}$}
 \end{array}\right.
\end{equation}
As a first guess we then calculate the $L_{\nu}(912)$ luminosity for each source  such that the total $\epsilon_{912}$ comoving emissivity in the volume is  
equal to the one derived by \citet{2015arXiv150202562G}, 
$2.5 \times 10^{24} \mathrm{erg/s/Hz/Mpc^3}$.
This results in a total ionising emissivity of $5.18\times10^{56}$ and $3.63\times10^{56}$ photons per seconds for the two sources. 
 In practice the ionising emissivity  as well as the corresponding 
fluctuations of the photo-ionisation rate  should  be 
  dominated by objects at the knee of the luminosity function 
  where there is currently no observational  constraint at this redshift. 
  The clustering of these sources and the enhanced mean-free path in their vincinity also plays a role.
   The two  sources in our toy model have a space density and luminosity roughly corresponding 
   to the region  where the   knee of the not yet probed QSO luminosity function at this redshift should lie.  
  We will explore  models with sources drawn from a realistic luminosity 
  function  and with a range of assumptions for their clustering in future work.

At each spatial position, we  compute the specific intensity of the ionising background between 1 and 4 Ryd 
by summing over the contribution from each bright source,
\begin{equation}
J(\mathbf{r},\nu)=\frac{1}{4 \pi}\sum_{i=1}^N \frac{L_i(\mathbf{r}_i,\nu)}{4\pi|\mathbf{r}_i-\mathbf{r}|^2}e^{\frac{-|\mathbf{r}_i-\mathbf{r}|}{ \lambda_{\mathrm{mfp}}^{912}}\left(\frac{\nu}{\nu_{912}}\right)^{-3(\beta-1)}},
\label{back_intensity}
\end{equation}

where, $\nu_{912}$ is the frequency at the HI ionising edge, and $\beta =1.3$ is
the slope of the HI column density distribution, which gives
the dependence of mean free path on frequency  (see also \citealt{2010ApJ...721.1448S} and \citealt{2013MNRAS.436.1023B}).
The sum in equation \ref{back_intensity} is performed for all sources within the periodically replicated simulation boxes 
(in practice, we included only the sources within the 100 Mpc/h box and its 26 directly neighboring periodic replications). 
We assume a mean free path $\lambda_{\mathrm{mfp}}^{912}$ as  extracted from our ATON `512-20-good\_h' full  radiative transfer simulation 
which has a value of $\lambda_{\mathrm{mfp}}^{912} \sim 35.09$ comoving Mpc/h at redshift $z\sim5.8$. 

The HI photoionisation rate due to bright sources is then computed as

\begin{equation}
 \Gamma(\mathbf{r})=4\pi \int_{\nu_{912}}^{4\nu_{912}}\frac{\mathrm{d}\nu}{h\nu}J(\mathbf{r},\nu)\sigma_{\mathrm{HI}}(\nu),
\label{gammaAGN}
\end{equation}

where $\sigma_{\mathrm{HI}}(\nu)$ is the photoionisation cross-section (calculated from the fit of \citealt{1997MNRAS.292...27H}).

We combine   the photoionisation rates  due to bright sources in our toy model as follows with the ionising UV background due to the much more numerous   galaxies driving reionisation in our ATON simulations
to calculate the expected effect of bright sources on the \lya opacity PDF.

\begin{itemize}
\item We randomly choose a line-of sight through  the $100 ({\rm Mpc/h})^3$ volume (along one of the principal axis)
         for which we have modelled  the contribution of bright sources to the photoionisation rate $
         \Gamma_{\mathrm{bright}}$.  
\item We  concatenate  three randomly  selected  line-of-sights from the `512-20-good\_l' full radiative  
        transfer simulation and place them along the line-of-sight through the bright source model and call the
        photoionisation rates  due to the sources in the radiative transfer simulation  $
       \Gamma_{\mathrm{gal}}$.
 \item We calculate the combined photoionisation rates  along  the line-of sight, $\Gamma_{\mathrm{gal \, + \, bright}}=
\beta_{\rm gal} \Gamma_{\mathrm{gal}}+\beta_{\rm bright} \Gamma_{\mathrm{bright}}$ .
 \item The value of $\beta_{\rm bright} $ is calculated  such  that for  chosen values of the number of bright sources and $\beta_{\rm gal}$  (the two additional free  parameter of our model once a mean free path has been chosen),  the mean value of the total photoionisation is that required to best match the \lya opacity PDF. 
Note that  when we calculate $\left< \Gamma_{\mathrm{gal \, + \, bright}} \right> $, we do so for  the entire 100 Mpc/h cube in  order to have the same value of $\beta_{\rm bright}$ in each line of sight.
 \end{itemize}

For  each  model for the spatial fluctuations of the photoionisation rates we  calculate 5000 mock \lya spectra 
for 50 Mpc/h chunks as in the observed sample of \cite{2015MNRAS.447.3402B} using the density, temperature, 
and peculiar velocity fields from the `512-20-good\_l' model. We have then varied  the number of bright sources and   $\beta_{\rm gal }$  with the aim of matching the observed \lya opacity PDF.
 In practice for a chosen number of bright sources we have started with 
choosing  $\beta_{\rm gal }=1$ and determined the value of $\beta_{\rm bright }$  required to match 
the mean photoionisation rate  in the `512-20-good\_h' model at  $z=5.8$ which is reasonable close 
to that required to mach the observed \lya opacity PDF. 
We have then further moderately rescaled   $\Gamma_{\mathrm{gal \, + \, bright}} $ to match the observed \lya opacity PDF as well as possible.

The model discussed in section \ref{PDF_teff}  for which we were able to obtain a good match had 
two bright sources, $\beta_{\rm gal } =1.15 $ and  $\beta_{\rm bright } =1.25$, 
but there will certainly be other parameter choices that will equally well reproduce the data.  In this model 
the galaxies driving reionisation and the bright sources responsible for the large scale fluctuations 
contribute about equally to the integrated ionising UV background. 
In figure \ref{gamma_toy_model}, we show the photoionisation rate $\Gamma_{\rm bright} $ in a slice of the  $100 ({\rm Mpc/h})^3$ volume for this model. 
For  comparison, we also show a map of a slice of $\Gamma_{\rm gal}$.
As expected our toy model produces of order unity fluctuations of $\Gamma$ on scales of $\ga$ 50 Mpc/h,
while the spatial fluctuations due to the much more abundant  galaxies in our radiative transfer simulations
occur on much smaller scales. 

It should, however, be kept in mind that the model presented  here neglects many aspects which will be important  
for an accurate modelling of the effect of bright sources. We have e.g. not self-consistently modelled the
effect of the bright sources on the mean-free path of ionising radiation. The large spatial extent   over which individual 
bright sources dominate also means that light-travel time effects become important if the sources are short-lived 
as is likely for bright starbursts and AGN.   The angular distribution of the emitted ionising radiation is also 
likely to be not isotropic. We should thus emphasise here that the results regarding the \lya opacity PDF 
based on this  simple  model  should just be seen as proof of concept.

\end{document}